\documentclass{aa}  
\usepackage{graphicx}
\usepackage{natbib}
\usepackage{txfonts}
\usepackage{gensymb}
\begin{document}

\title{The outermost stellar halo of NGC 5128 (Centaurus A):  Radial structure\thanks{Photometry data tables are only available in electronic form
at the CDS via anonymous ftp to cdsarc.u-strasbg.fr (130.79.128.5)
or via http://cdsweb.u-strasbg.fr/cgi-bin/qcat?J/A+A/}}

\author{M.~Rejkuba\inst{\ref{inst1}}
    \and W.~E.~Harris\inst{\ref{inst2}}
    \and L.~Greggio\inst{\ref{inst3}}
    \and D.~Crnojevi\'c\inst{\ref{inst4}}
    \and G.~L.~H.~Harris\inst{\ref{inst5}}
          }

\institute{European Southern Observatory, Karl-Schwarzschild Strasse 2, D-85748 Garching, Germany; 
              \email{mrejkuba@eso.org}\label{inst1}
         \and
                 Department of Physics and Astronomy, McMaster University, Hamilton ON L8S 4M1, Canada\label{inst2}
          \and 
             INAF, Osservatorio Astronomico di Padova, Vicolo dell'Osservatorio 5, 35122 Padova, Italy\label{inst3}
              \and 
          University of Tampa,  Department of Chemistry, Biochemistry, and Physics, 401 West Kennedy Boulevard, Tampa, FL 33606, USA\label{inst4}
          \and
                 Department of Physics and Astronomy, University of Waterloo, Waterloo ON N2L 3G1, Canada\label{inst5}
          }

\date{Received date; accepted date}

  \abstract
   {The extended stellar halos of galaxies contain important clues for investigating their 
   assembly history and evolution. }
   {We investigate the resolved stellar content and the extended halo of NGC~5128 as a function of galactocentric distance, and trace the halo outward to its currently detectable limits.}
   {We used Hubble Space Telescope images obtained with the WFPC2, ACS, and WFC3 cameras equipped with $F606W$ and $F814W$ filters to resolve individual red giant branch (RGB) stars in 28 independent pointings across the halo of NGC~5128. The stellar halo analysis for 14 of these pointings is presented here for the first time.
   Star counts from deep $VI$ color-magnitude diagrams reaching at least 1.5 mag below the tip of the RGB are used to derive the surface density distribution of the halo. The contamination by Milky Way stars is assessed with a new control field, with models, and by combining optical and near-IR photometry.}
   {We present a new calibration of the WFC3 F606W+F814W photometry to the ground-based $VI$ photometric system. The photometry shows that the stellar halo of NGC~5128 is dominated by old RGB stars that are present in all fields. The V-band surface brightness of fields changes from 23 to 32 mag~arcsec$^{-2}$ between the innermost field only 8.3~kpc from the galaxy center to our outermost halo fields, which are located 140~kpc away from the center along the major axis and 92~kpc along the minor axis.
   Within the inner $\sim 30$~kpc, we also find evidence for a $2-3$~Gyr old population traced by asymptotic giant branch stars that are brighter than the tip of the RGB. This population contributes only up to 10\% in total stellar mass if it is  2~Gyr old, but a larger fraction of $30-40$\% is required if its age is 3~Gyr. The stellar surface density profile is well fit by a classic r$^{1/4}$ curve or a simple power-law form $\sim r^{-3.1}$ over the full radial range, with no obvious break in the slope, but with large field-to-field scatter.  The ellipticity measured from integrated-light photometry in the inner parts, $e=(b/a) = 0.77$, flattens to $e=0.54 \pm 0.02$ beyond 30~kpc. Considering the flattening of the outer halo, the projection of the elliptical isophote on the semimajor axis for our most distant field reaches nearly 30 effective radii.
   }
   {}
   
\keywords{Galaxies: elliptical and lenticular, cD -- 
             galaxies: individual: NGC~5128 -- 
             Galaxies: stellar content
               }

\maketitle
 
%

\section{Introduction}
\label{sec:intro}

We have long known that the 
halo of our own Milky Way galaxy extends over many kiloparsecs and that it is principally populated
with metal-poor stars \citep{ELS62, searle+zinn78, freeman1987}. The high level of inhomogeneity, the clear presence of stellar streams
and substructure, and the possible dual nature of the halos of Milky Way and M31
are among the more recent results that have changed our understanding of galaxy halo assembly 
\citep[e.g.,][]{ferguson+02,belokurov+06,carollo+07,bell+08,
sesar+11,drake+13,ibata+14}. 

A major difference for studies of stellar halos around nearby galaxies beyond the Local Group was made by the high resolution and high sensitivity of the Hubble Space Telescope (HST) cameras, which can resolve and measure individual halo stars routinely in galaxies at distances out to about 10~Mpc \citep{harris+07a,rejkuba+11,peacock+15,monachesi+16,cohen+20}.  With special efforts to accumulate longer exposure times, galaxies out to the distance of the Virgo system are within reach \citep{williams+07,bird+10}. 
Direct starcounts of RGB stars make a highly
effective route to tracing the halo population outward to levels of
equivalent surface brightness that are difficult to achieve by other means
\citep[e.g.,][in addition to the papers cited above]{pritchet_vandenbergh1994,harris+07b,gilbert+2012,harmsen+2017}.
Integrated-light studies can be quite effective at identifying the spatial distribution and structure of the halo \citep{mihos+13, duc+15, merritt+16, iodice+16, iodice+19, beilek+20}, in spite of challenging photometry at extremely low surface brightness levels (see \citealt{mihos19}
for a recent review of advances and challenges related to deep imaging of diffuse light around galaxies). However, integrated-light studies do not allow an unambiguous description of the properties of the stellar populations because of the age-metallicity degeneracy \citep[see, e.g.,][]{mihos+13}, and furthermore, they do not give information on the \textit{\textup{distributions}} of age and metallicity.  
These underlying distributions
can be addressed by analyzing the color-magnitude diagram (CMD) of resolved stars. 

Even with the HST, only a few large early-type galaxies (ETGs) are especially amenable to resolved stellar population studies.  Of these,
NGC 5128 (often referred to by its well-known radio source designation
Centaurus A or Cen A) has a special place: At a distance of 3.8~Mpc \citep{rejkuba04,
harris+10}, it is by far the nearest easily observable giant ETG.
Because of its proximity, NGC 5128 was the first giant ETG 
to have its stellar halo
resolved into individual red giant branch (RGB) stars 
\citep{soria+96}, and it has since been observed with the HST within several programs. It has also been imaged under very good seeing conditions from the ground. 

Several earlier studies concentrated on the visible star formation in the inner halo
\citep{fassett+graham00,mould+00,rejkuba+01,rejkuba+02,graham+fassett02,peng+02,crockett+12}, 
which is associated with the so-called inner and outer filaments \citep{blanco+75, santoro+15}  
and is possibly triggered by the radio jet \citep{oosterloo+morganti05}. Although visibly quite prominent, these features contain only a small fraction of the halo mass 
\citep{rejkuba+04,oosterloo+morganti05}, have a low star formation efficiency \citep{salome+16}, 
and are confined to a region extending out to$\simeq 35$ kpc along the northeastern major axis \citep{neff+14}. Within a radius of $\sim 20$~kpc lie also well-known shells around the galaxy \citep{malin+83,peng+02} that were likely formed during a past
accretion of a companion galaxy that deposited the gas and dust that has long been noted as a characteristic of NGC~5128 \citep{charmandaris+00}. \citet{struve+10} pointed out that the  H\i\, fraction in Centaurus~A  ($M_{\mathrm{HI}}/L_B = 0.01$) is rather low for an ETG, suggesting that the most recent accretion involved a relatively small (Small Magellanic Cloud like) galaxy about $1.6 - 3.2 \times 10^8$~yr ago. An alternative major-merger origin was proposed by numerical simulations that reproduced some selected properties of  NGC~5128  \citep{bekki+peng06,wang+20}. Setting this in a broader context: 
\citet{tal+09} found that 73\% 
of nearby luminous elliptical galaxies show tidal disturbance signatures in their stellar
bodies; similar results are evident from the MATLAS survey \citep{duc+15}. Thus
NGC~5128, far from being ``peculiar'', is a quite typical giant elliptical \citep[see also the review of][]{harris10}.

A systematic survey of the stellar halo properties in NGC~5128 was performed with HST imaging
that reached at least 1.5 mag below the RGB tip (TRGB), deep enough to sample its full metallicity distribution. This survey started with the
works of G.~Harris and collaborators over 20 years ago. Inner- and mid-halo fields were
observed with the WFPC2 camera at 8, 21, and 31 kpc distance from the center
\citep{harris+99,harris+harris00,harris+harris02}, establishing for the first time a (then) surprisingly broad
metallicity distribution of the halo RGB stars, with a peak metallicity
close to solar in the innermost field and only mildly subsolar in the other two
fields.  Low-metallicity stars reaching [Fe/H] $\simeq -2$~dex are present in all these
fields, but are very much in the minority.

After the HST servicing mission 3B and the installation of the Advanced Camera 
for Surveys (ACS), we carried out a much deeper photometric probe in
a field 38 kpc SSE of the center, with sufficiently faint limits to detect the core helium burning stars
located in the red clump \citep{rejkuba+05}. This deep view of the halo not
only further confirmed the metal-rich nature of the halo stars, but also permitted a
quantitative estimate of the halo age distribution: 70-80\% of the stars formed $12\pm 1$~Gyr ago, and the
remaining 20-30\% population is best fit with 2-4 Gyr old 
models \citep[][henceforward R11]{rejkuba+11}.
Remarkably, these deep data indicate that the full metallicity range of the
models ($Z=0.0001-0.04$) combined with old ages needs to be used to reproduce the
colors of the reddest RGB stars in this field. 

The last Hubble servicing mission 4 enabled imaging with a wider field of 
view by combining the newly installed Wide Field Camera 3 (WFC3) in parallel with ACS. 
We designed a program to map five pairs of new locations in
the outer halo of NGC~5128, two along the minor and three along the northeastern major
axis, with the intention of exploring the extent of the halo and the properties of the stars located
in its most extreme regions. The first results from this program were summarized briefly in
\citet[][Paper I]{rejkuba+14}. 
We found a transition from a metal-rich inner galaxy
to a lower-metallicity outer halo, with a shallow metallicity gradient 
and  hints
of possible substructures in the outer halo based on metallicity and number density 
variations in neighboring parallel fields. The shape of
the halo was found to be elongated (roughly consistent with the inner halo), 
with an excess of stars along the major axis 
above the $r^{1/4}$ law fit to the star counts from earlier HST studies. Most
remarkably, halo RGB stars are still present in fields as far out as 140~kpc (25 effective radii
$R_e$) along the major axis, and 90~kpc (16 $R_e$) 
along the minor axis.  In short, no clear `end' to this galaxy halo has been found.

Complementing the narrow pencil-beam studies conducted with the HST, 
the extended halo of NGC 5128 was also surveyed from the 
ground with the VIMOS\footnote{VIsible MultiObject Spectrograph (VIMOS) was mounted on the Unit Telescope 3 (UT3) of the Very Large Telescope (VLT) at the European Southern Observatory (ESO) Paranal Observatory and it included imaging in addition to its primary spectroscopic modes.} optical imager on the 8m ESO VLT 
\citep{crnojevic+13, bird+15}. The Magellan Megacam imager at Las Campanas Observatory was used to observe NGC 5128
\citep{crnojevic+16} as part of the Panoramic Imaging Survey 
of Centaurus and Sculptor \citep[PISCeS;][]{crnojevic+16iaus}. 
These wide surveys uncovered a vast amount of substructure, including several dwarfs in the process of being accreted. This process creates overdense regions in the halo. However, the intermediate Galactic latitude of NGC 5128 unfortunately means that large numbers of  foreground stars are present, which adds to unresolved background galaxies. 
 When the field contamination issue is combined with uneven completeness due to observations taken under a range of observing conditions, 
studying its halo properties based on RGB star counts is challenging from the ground. 
We believe that the best way forward is
to combine  the strengths of the wide-area surveys that provide a global view of the halo and substructure with the higher resolution and deeper observations from space that enable detailed investigation of 
its stellar composition. 

Since our last HST-based study (Paper I), new ACS and WFC3 imaging was secured that primarily focused on confirming the newly discovered dwarf galaxies and substructures \citep{crnojevic+19}, but also added further parallel pointings in the halo of NGC~5128.  Armed with the information from the recent wide-area PISCeS survey of the Cen A halo, 
we assembled all HST observations taken so far to present a homogeneous analysis of the radial structure of the halo, its metallicity distribution, and metallicity gradients across a vast area of the halo of NGC 5128.   


\begin{figure*}
\centering
\resizebox{\hsize}{!}{
\includegraphics{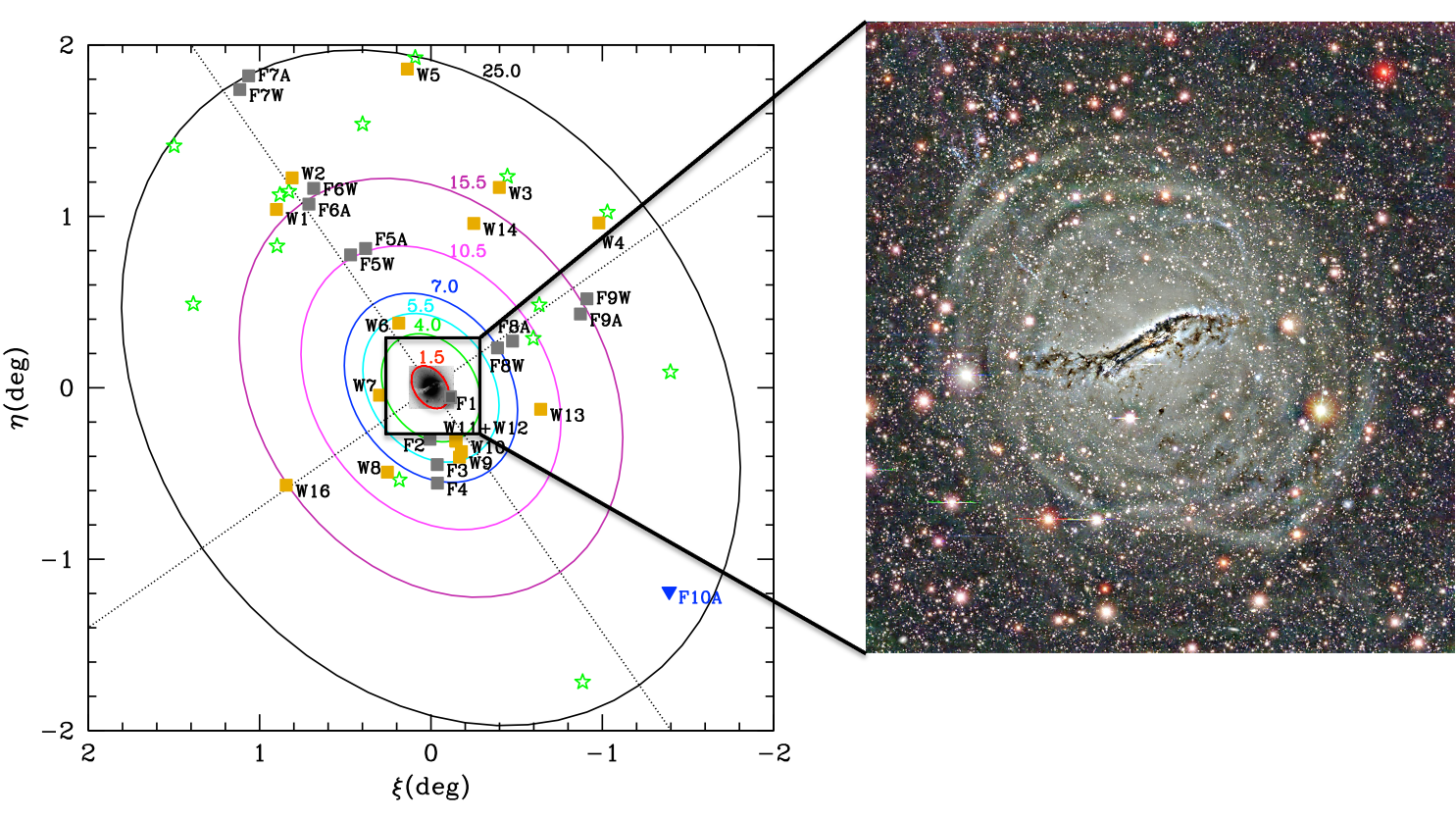}}
\caption{Distribution of the fields imaged with the HST in F606W and F814W filters
as listed in Table~\ref{tab:HSTobslog} relative to the
center of NGC~5128 projected on the plane of the sky. Gray squares are fields F1-F9, orange squares are for WFC3 parallels W1-W16 from GO13856, and the blue inverted triangle is for the ACS parallel from GO15426. 
North is at the top and east at the left. The halo isophotal major
axis is oriented 35\degree \ east of north \citep[counterclockwise;][]{dufour+79}.  The elliptical contours 
have an axis ratio of 0.77 as determined from the inner halo \citep{dufour+79} and are plotted for 
1.5, 4, 5.5, 7, 10.5, 15.5, and 25 $R_e$ distance (as indicated). The green stars 
are dwarf galaxies in the Cen A group as listed in Table 4 in \citet{mueller+19}. Most of them have recently confirmed Cen A group membership based on the primary pointings from GO13856 \citep{crnojevic+19}. 
The large black square in the center indicates the relative size of the $35'\times 35'$ ($38.6 \times 38.6$~kpc$^2$)
unsharp-masked image of the shells in the central parts of the galaxy shown in
the right panel \citep[][image credit: Eric Peng (JHU), Holland Ford (JHU/STScI), Ken Freeman
(ANU), Rick White (STScI), and NOAO/AURA/NSF]{peng+02}.}
\label{fig:fieldsrel}
\end{figure*}


In this paper we describe in more detail the data and photometric
measurements done for the Cycle 20 observations, 
for which initial results were presented in Paper I.
We add 15 more halo fields that were obtained as Parallel pointings within the more recent 
Cycle 22 and Cycle 25 HST programs (PI: Crnojevi\'c), essentially doubling the database.
We also carry out a homogeneous analysis of all these halo locations. These parallel fields were only used as control fields for the dwarf galaxy CMD analysis by \citet{crnojevic+19} and were not analyzed in terms of their contribution to the overall stellar halo properties before.
Furthermore, we discuss the foreground and background field contamination using several methods: a control field, Milky Way models, and a combination of optical and near-IR photometry. We finally adopt a statistical decontamination of our CMDs based on the control field. 
From the newly calibrated and decontaminated CMDs, we derive a new self-consistent surface-density profile extending outward to the currently detectable limits of this galaxy's giant halo and measure the stellar halo ellipticity.
We discuss the possible presence of intermediate-age asymptotic giant branch (AGB) stars in the halo by
analyzing the numbers of stars populating the CMDs just above the RGB tip.

Throughout the paper, we adopt the distance to NGC 5128 of 3.82~Mpc  (distance modulus $(m-M)_0 = 27.91$ mag), which is based on average distances from four well-established distance determination methods: the Cepheid period-luminosity relation, TRGB, the planetary nebulae luminosity function, and the Mira variables period-luminosity relation (see \citealt{harris+10} for detailed discussion of the distance determinations). Based on the intrinsic brightness of the RGB tip $M_I(TRGB)=-4.05$ \citep{rizzi+07,freedman+20}, the TRGB in NGC~5128 fields is found at $I_0 = 23.86$ mag.


\begin{table*}
\caption{HST observations of the NGC 5128 halo fields. 
}
\label{tab:HSTobslog}
\centering
\begin{tabular}{lclrrlllrrrr}
\hline\hline
\multicolumn{1}{c}{Field} &  
\multicolumn{1}{c}{Date} &  
\multicolumn{1}{c}{Prg.}  &
\multicolumn{1}{c}{Exptime}  &
\multicolumn{1}{c}{Exptime}  &
\multicolumn{1}{c}{HST}  &
\multicolumn{1}{c}{RA}      &
\multicolumn{1}{c}{DEC}     &
\multicolumn{1}{c}{E(B$-$V)}     &
\multicolumn{1}{c}{$R$}     &
\multicolumn{1}{c}{$\theta_a$} &
\multicolumn{1}{c}{$a/R_e$} \\
\multicolumn{1}{c}{ID$^1$}           &  
\multicolumn{1}{c}{yyyy-mm} &  
\multicolumn{1}{c}{ID}      &
\multicolumn{1}{c}{F814W}  &
\multicolumn{1}{c}{F606W}  &
\multicolumn{1}{c}{Camera}  &
\multicolumn{1}{c}{J2000}      &
\multicolumn{1}{c}{J2000}     &
\multicolumn{1}{c}{(mag)}     &
\multicolumn{1}{c}{(kpc)}     & 
\multicolumn{1}{c}{(deg)} \\
\multicolumn{1}{c}{}           &  
\multicolumn{1}{c}{} &  
\multicolumn{1}{c}{}    &
\multicolumn{1}{c}{(sec)}  &
\multicolumn{1}{c}{(sec)}  &
\multicolumn{1}{c}{}  &
\multicolumn{1}{c}{hh:mm:ss}      &
\multicolumn{1}{c}{dd:mm:ss}     &
\multicolumn{1}{c}{}     &
\multicolumn{1}{c}{}     &
\multicolumn{1}{c}{}     \\
\hline
F1    & 1999-06   &  8195 &12100 & 17500 &   WFPC2/PC     & 13:24:51 & -43:04:33  &  0.121 &   8.3  & 152.0 & 1.8  \\
F2    & 1997-08   &  5905 &12800 & 12800 &   WFPC2/WF     & 13:25:29 & -43:19:09  &  0.115 &  20.0  & -144.2 & 4.8  \\
F3    & 1999-07  &  8195 &17500 & 17800 &   WFPC2/WF+PC & 13:25:16 & -43:28:01  &  0.123 &  29.9  &  -149.5 & 6.8  \\
F4    & 2002-07  &  9373 &30880 & 30880 &   ACS/WFC       & 13:25:15 & -43:34:30  &  0.123 &  37.1  &  -149.0 & 8.4  \\
F5A   & 2013-05  &  12964 &2137  & 2270  &   ACS/WFC      & 13:27:32 & -42:12:24  &  0.115 &  59.8  &  10.0 & 11.0  \\
F5W   & 2013-05  &  12964 &2376  & 2496  &   WFC3/UVIS    & 13:28:01 & -42:14:37  &  0.113 &  60.5  &  3.9 & 10.8  \\
F6A   & 2013-01  &  12964 &2137  & 2270  &   ACS/WFC      & 13:29:19 & -41:56:51  &  0.141 &  85.7  &  1.7 & 15.2  \\
F6W   & 2013-01  &  12964 &2376  & 2496  &   WFC3/UVIS    & 13:29:10 & -41:51:21  &  0.139 &  90.0  &  4.8 & 16.0  \\
F7A   & 2013-06  &  12964 &2137  & 2270  &   ACS/WFC      & 13:31:12 & -41:11:57  &  0.091 & 140.5  &  5.0 & 25.0  \\
F7W   & 2013-06  &  12964 &2376  & 2496  &   WFC3/UVIS    & 13:31:29 & -41:16:45  &  0.093 & 137.8  &  2.7 & 24.4  \\
F8A   & 2013-05  &  12964 &2137  & 2270  &   ACS/WFC      & 13:22:52 & -42:44:49  &  0.139 &  36.5  &  95.1 & 11.9  \\
F8W   & 2013-05  &  12964 &2376  & 2496  &   WFC3/UVIS    & 13:23:20 & -42:47:10  &  0.143 &  30.3  &  94.1 & 9.9  \\
F9A   & 2013-01  &  12964 &2137  & 2270  &   ACS/WFC      & 13:20:42 & -42:35:19  &  0.158 &  64.9  &  98.7 & 21.1  \\
F9W   & 2013-01  &  12964 &2376  & 2496  &   WFC3/UVIS    & 13:20:30 & -42:30:00  &  0.144 &  69.9  &  95.2 & 22.8  \\
W1    & 2015-08  & 13856  &1278  & 1210  &  WFC3/UVIS     & 13:30:21 & -41:58:43 & 0.110 & 91.8 & -5.7 & 16.4  \\ 
W2    & 2016-01 & 13856  &1278  & 1210  &  WFC3/UVIS     &13:29:51  & -41:47:41 & 0.112 & 97.9 & 1.8 & 17.3  \\ 
W3    & 2015-06 & 13856  &1278  & 1210  &  WFC3/UVIS     &13:23:18  & -41:51:02 & 0.110 & 82.3 & 53.7 & 23.4 \\
W4    & 2015-06 & 13856  &1278  & 1210  &  WFC3/UVIS     &13:20:08  & -42:03:27 & 0.107 & 91.6 & 80.4 & 29.6  \\
W5    & 2015-06 & 13856  &1278  & 1210  &  WFC3/UVIS     &13:26:12 & -41:09:35 & 0.094 & 124.3 & 30.84 & 28.2  \\
W6    & 2016-01 & 13856  &1278  & 1210  &  WFC3/UVIS     &13:26:29 & -42:38:31 & 0.096 & 28.1 & 8.6 & 5.1  \\
W7   &  2015-06 & 13856  &1278  & 1210  &  WFC3/UVIS     &13:27:07 & -43:03:44 & 0.089 & 20.2 & -63.1 & 6.2 \\
W8   &  2015-04 & 13856  &1278  & 1210  &  WFC3/UVIS     &13:26:51  & -43:30:40  & 0.099 & 36.9 & -117.7 & 11.1  \\
W9    & 2015-04 & 13856  &1278  & 1210  &  WFC3/UVIS     & 13:24:33 & -43:25:25 & 0.114 & 29.1 & -167.4 & 5.5  \\
W10   & 2015-04 & 13856  &1278  & 1210  &  WFC3/UVIS     & 13:24:29  & -43:23:30 & 0.116 & 27.5 & -170.6 & 5.0  \\
W11   & 2015-06 & 13856  &1278  & 1210  &  WFC3/UVIS     & 13:24:39 &  -43:19:50 & 0.115 & 23.0 & -170.4 &4.2  \\
W12   & 2015-04 & 13856  &1278  & 1210  &  WFC3/UVIS     & 13:24:40 &  -43:19:37 & 0.115 & 22.7 & -170.2 & 4.2  \\
W13   & 2015-06 & 13856  &1278  & 1210  &  WFC3/UVIS     & 13:21:57 & -43:08:40 & 0.130 & 43.5 & 136.1 & 11.4  \\
W14   & 2016-01 & 13856  &1278  & 1210  &  WFC3/UVIS     & 13:24:06 & -42:03:37 & 0.106 & 66.1 & 49.5 & 18.1  \\
W16   & 2015-06 & 13856  &1278  & 1210  &  WFC3/UVIS     & 13:30:06 & -43:35:15 & 0.117 & 67.8 & -88.8 & 22.3  \\
\\
F10A$^2$   & 2018-04 & 15426 & 2228 & 2379 & ACS/WFC & 13:17:47 & -44:12:24 & 0.096 & 121.8 & 165.2 & 23.4 \\
\hline
\end{tabular}
\tablefoot{
Columns RA and DEC report the center of the field coordinates, E(B$-$V) is the reddening value toward the field center based on \citet{schlegel+98} reddening maps, and $R$ is the projected radial distance in kpc from the center of the galaxy. In the last two columns, $\theta_a$ is the azimuthal angle of the field location measured clockwise from the isophotal major axis (see Fig.~\ref{fig:fieldsrel}), while $a$ is the semimajor axis of the elliptical isophote that goes through the field (assuming e=0.54 ellipticity) in units of $R_e=305" = 5.6$~kpc. \\
\tablefoottext{1}{Our numbering of the Wxx fields in the first column (Field ID) follows the MAST Archive listing, in which there
is no ``W15''.\\}
\tablefoottext{2}{The field F10A listed at the bottom is used as a background control field (see Sect.~\ref{sec:decontstat}).}
}
\end{table*}

\section{Data}

\subsection{HST imaging observations}
\label{sec:HSTimaging}

In Table~\ref{tab:HSTobslog} we list the basic information about all HST fields used 
in this study.\footnote{Additional images for other
fields in NGC~5128 exist in the
HST archive, but they are unsuitable for a homogeneous analysis because of one of the following reasons: (i) located at the center of known overdensities \citep{crnojevic+19}, (ii) are not deep enough, or (iii) were obtained with different filters, or only in one filter.} Their distribution on the sky, in a coordinate system centered on NGC 5128, is shown in Fig.~\ref{fig:fieldsrel}. 
In the right panel we also show the unsharp-masked image from \citet{peng+02}.  The comparison emphasizes the wide span
of the halo probed by our target fields, most of which lie far beyond the inner regions (i.e., the inner square in Fig.~\ref{fig:fieldsrel})
containing arcs and shells, dust, and star-forming regions that are remnants of the recent merger episode(s).
Concerns about contamination of our target fields from these factors are thus mitigated.  
By contrast, we need to be aware of possible contamination from the presence of outer-halo dwarf
satellites and substructure \citep{crnojevic+16}. We show therefore in Fig.~\ref{fig:fieldsrel} the location of known dwarf satellites of Cen~A within the area covered by our HST fields. Having effective radii between $\sim 150-600$~pc, these dwarfs typically cover up to $ \sim 1'$ on sky\footnote{1' at the distance of Cen A corresponds to $\sim 1.1$~kpc.} and thus do not contribute stars to adjacent parallel pointings. One exception is the CenA-MM-Dw3 stream \citep{crnojevic+19} that crosses one of our fields. This is further discussed below.

The observations of the inner WFPC2 fields F1-F3 taken within HST Cycles 5 and 8 were described 
in detail by \citet{harris+99} and \citet{harris+harris00, harris+harris02}, while field F4 is our deepest probe into the halo population,  
where the ACS photometry reached the core-helium burning stars \citep{rejkuba+05,rejkuba+11}. This deep field received 12 orbit-long exposures for each of the two filters. The fields 
F5-F9 are from our Cycle 20 data (Paper I) and have two entries each, one for the WFC3 camera (e.g., F5W), and the other for the ACS (e.g., F5A) images taken in parallel mode. The Cycle 20 observations were organized such that two orbits for each target were placed within a spacecraft visit to obtain the images in two filters at the same field orientation.  Within each orbit, a set of three dithered images using the WFC3-UVIS-DITHER-LINE-3PT pattern were taken for each filter. 

Fields W1 -- W16 are the WFC3 
observations from the Cycle 22 program GO13856 (PI: Crnojevi\'c).  In this  program, ACS fields were placed on faint dwarf satellites \citep{crnojevic+19}, while the WFC3 data were the parallels located on blank halo fields
adjacent to them.  The two cameras are separated by more than $5'$ in the focal plane of the HST, corresponding
to a linear separation of $\sim 5.5$ kpc center to center at the distance of NGC 5128. This distance is much larger than
the physical dimensions of any of the dwarf satellites, therefore the W1 -- W16 pointings are expected to sample pure field halo
populations.  Fields W11 and W12 heavily overlap in location due to the closeness of the primary ACS targets, and there is no W15 field (see Table \ref{tab:HSTobslog}). The Wxx set therefore really just covers 14 independent pointings. 

Finally, we added one more field (F10A, listed in the last row of Table \ref{tab:HSTobslog}).
This is the parallel image from program GO15426 (PI Crnojevi\'c) taken with the ACS, 
located on a remote halo field $\sim 120$ kpc to the southwest.  As we describe below,
it provides us with what is likely to be the best available control field for gauging the field
contamination.

For all the targets listed in Table \ref{tab:HSTobslog}, the filters were F606W and F814W, which transform well
into $(V, I)$ and which provide a color index $(V-I)$ that is reasonably sensitive to metallicity
for old RGB stars over the entire range from [Fe/H] $\simeq -2$~dex to above-solar abundance.
The "Exptime" columns in Table~\ref{tab:HSTobslog} report the total exposure time per filter
(for F5-F9, these are essentially full one-orbit exposures, while for W1-W16, they are half-orbit 
exposures, leading to slightly shallower CMDs).  The camera(s) that were used determine the total field of view (FOV) per field:  
for F1 (WFPC2/PC1), this is as small as 0.3403 arcmin$^2$, F2 covers 5.33 arcmin$^2$, and F3, which combines all four WFPC2 detectors, reaches 5.674 arcmin$^2$.  For the rest,  
the ACS camera FOV is 11.33 arcmin$^2$ , and WFC3 covers 7.29 arcmin$^2$. 

In total the target fields span almost 4 degrees across the sky. We therefore expect 
differences in Galactic (Milky Way) foreground extinction from one field to the next that are large enough to call for individual
correction. The adopted Galactic extinctions toward each field are listed in Table \ref{tab:HSTobslog} and are computed from the \citet{schlegel+98} reddening value E(B$-$V) adopting the \citet{schlafly+finkbeiner11} recalibration\footnote{source: http://irsa.ipac.caltech.edu/applications/DUST/} and $A_V=2.79*E(B-V), A_I=1.55*E(B-V)$. 

Table~\ref{tab:HSTobslog} also gives the on-sky projected radial distance $R$ in kiloparsecs from the center of NGC 5128\footnote{RA$_0$=13:25:27.6, DEC$_0$=-43:01:09 from NED} assuming an intrinsic distance modulus 
$(m-M)_0 =  27.91$ \citep{harris+10}. The position angle $\theta_a$ is the azimuthal angle (measured north of east, or clockwise in the figure) 
from the major axis $a$ 
(where the ellipticity is $e = (b/a) = 0.54$, see Sect.~\ref{sec:AGB} below).
Here $a$ is listed in units 
of $R_e$, where we adopt $R_e = 305\arcsec = 5.6$ kpc from \citet{dufour+79} as determined from the integrated-light
profile of the inner spheroid.

\subsection{Photometry}

All observations used the same filters ($F606W, F814W$) and the photometric data reduction procedures were the same. We started from the pipeline processed \emph{*.drc} images downloaded from the HST archive that are the multidrizzled combinations of the individual exposures including CTE corrections. 
While the stars in the target halo fields are quite uncrowded in any absolute sense and aperture photometry would be possible for them, we preferred to run the 
\emph{daophot} and \emph{allstar} suite of photometric codes in IRAF \citep{stetson87} to perform the photometry via PSF fitting.  
In addition to a homogeneous procedure applied to all fields, this
gives us the advantage of having stellar PSF-fitting parameters that can be used to
distinguish objectively between slightly resolved background galaxies and point sources. 

In each field, a master image consisting of all exposures in both filters was constructed to provide
the deepest possible source for object detection.
SourceExtractor \citep{bertin_arnouts1996} was used on the master image
to detect candidate objects and perform a preliminary culling out of
nonstellar objects (half-light radii $r_{1/2} < 1.0$ or $> 1.6$ px were rejected).  From there, \emph{daophot}
was used to carry out small-aperture (r = 2 px) photometry, construction of a PSF for each filter, and
then final photometry with \emph{allstar}.  The independently determined PSFs on each field
were built from typically 30 to 70 individual bright uncrowded stars.
Although these 
proved to be quite consistent with each other, in the end, to improve internal
consistency, exactly the same set of $(F606W,F814W)$ PSFs picked from the highest S/N cases for a given camera and filter were used
on all the fields.  Further culling of the starlists was done by rejecting any objects with 
\emph{allstar} parameters $\chi > 2$ or $err > 0.2$ mag in either filter. Because crowding is 
not a factor, any such rejected objects are almost always nonstellar.
Empirically derived aperture corrections to r = 10 px were added to
the \emph{allstar} measured magnitudes, and the large-aperture data were then converted into
filter magnitudes $(F606W, F814W)$ with filter zeropoints from the ACS or WFC3 webpages.
These were finally converted into standard $(V,I)$ with the linear color transformations noted below.

\subsection{Calibrations for the ACS and WFC3 photometry}

Our photometric calibration for the ACS fields follows the prescription in \citet{sirianni+05}
and is on the VEGAMAG system. To bring
the instrumental magnitude measurements to the HST system, we used the zeropoint calculator 
published on the HST web pages.
We applied the following calibration equations:
\begin{equation} 
F606W = -2.5\times \log{\frac{f606w_{inst}}{\mathrm{exptime}}} + zpt_{F606W} + apcor_{F606W} 
\end{equation}
\begin{equation}
F814W = -2.5\times \log{\frac{f814w_{inst}}{\mathrm{exptime}}} + zpt_{F814W} + apcor_{F814W} 
,\end{equation}
where $zpt_{\mathrm{filter}}$ is the VEGAMAG zeropoint for the given filter, and $apcor_{\mathrm{filter}}$ is the corresponding aperture correction from our PSF measurement to a 10 px aperture, and then from 10 px 
to infinite radius \citep{bohlin12}.  For the ACS data, 
$zpt_{F606W}=26.407$ and $zpt_{F814W}=25.523$.

For the WFC3 camera, the aperture correction to 10 px  = $0\farcs4$ radius was made, after which filter zeropoints can be applied that already include the step to infinite radius. We adopted $zpt_{F606W}=25.8843$, $zpt_{F814W}=24.5730$.

In Paper I we used the WFC3 zeropoints from the HST webpages to convert the $(F606W, F814W)$ magnitudes into $(V,I)$ , but out of temporary necessity, we applied the color terms ($c_1$) for the same filters from the ACS/WFC detector. Since then, Sahu et al. (2014) provided additional data for WFC3/UVIS, listing the zeropoint differences $(I-F814W), (V-F606W)$ for stars over a range of blackbody 
temperatures and spectral types.  By plotting these differences versus $(V-I)$,
we reconstructed the color terms in the Vegamag system for the transformations.  As with ACS, 
the slope of the color term $c_1$ is very small for $I$, but for $V$ , it appears to be slightly shallower than for the ACS. 
The transformations for WFC3 that we adopt here are 
\begin{equation}
(V-I) = 1.1396 \times (F606W-F814W)
\end{equation}
\begin{equation}
V = F606W + 0.1545 \times (V-I)
\end{equation}
\begin{equation}
I = F814W + 0.032 \times (V-I)
.\end{equation}

Empirical transformations of the standard WFC3 filters into $BVI$ have also
been derived by \citet{harris2018} from combined ACS and WFC3 photometry in a
47 Tucanae standard field; encouragingly, for F606W and F814W, these are quite similar to those given above.  The effect of adopting the revised transformations above was to make the $(V-I)$ colors from WFC3 bluer than in Paper I by typically $\Delta(V-I) \simeq 0.1$ mag.  This
shift brings the CMDs for ACS and WFC3 into closer agreement.

\subsection{Completeness and photometric error analysis}

\begin{figure}
\resizebox{\hsize}{!}{
\includegraphics{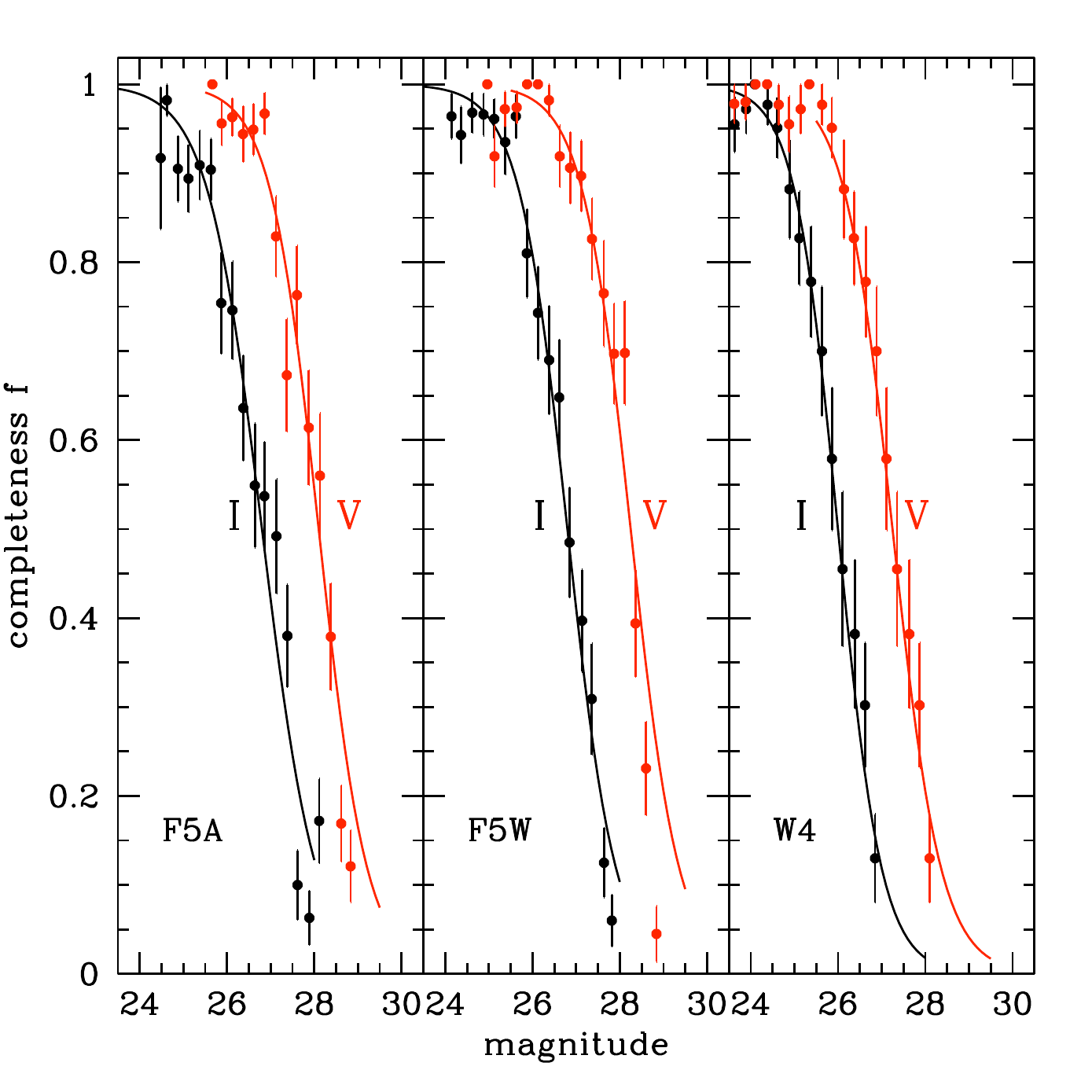}
}
\caption{Samples of \emph{daophot/addstar} experiments for three different
fields in our study.  The completeness fraction $f$ is plotted vs.
V or I magnitude in each panel.  The measured $f-$values per 0.25-magnitude bin 
are shown as the black (I) or red (V) points with error bars.  The solid lines
(black for I, red for V) show the smooth fitted curves of the form 
$f = 1 / (1 + e^{\alpha (m-m_0)})$ 
described in the text.
}
\label{fig:completeness}
\end{figure}

\begin{figure}
\resizebox{\hsize}{!}{
\includegraphics{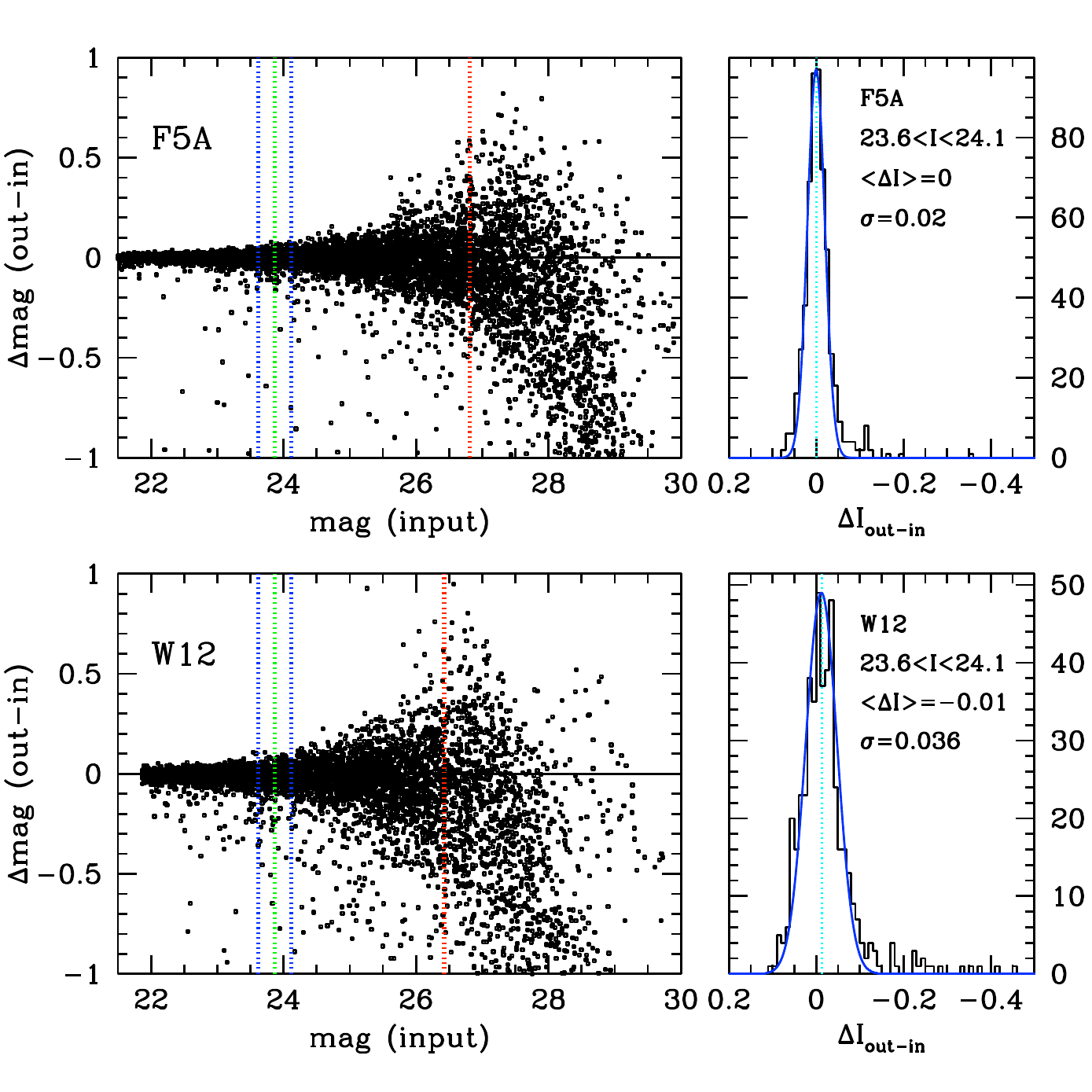}}
\caption{Sample of photometric measurement precision for Field F5A (top) and W12 (bottom panels) for the I band.  The magnitude difference $\Delta(m)$ (measured -- input) for 10000 input artificial
stars is plotted against input magnitude for F5A (upper left) and W12 (lower left panel). The 50\% detection completeness
levels are marked with the vertical dotted red lines. The dotted green line is the magnitude of the TRGB. The blue lines indicate $\pm 0.25$mag around the TRGB. In the right panels we plot the distributions of magnitude differences (measured -- input) for stars within $\pm 0.25$~mag of the TRGB, binned to 0.01 mag for F5A (upper right) and W12 (lower right panel). The Gaussian curves plotted over the magnitude difference histograms have the mean and sigma as indicated in the panels. 
}
\label{fig:delmag}
\end{figure}

\begin{table}
\caption{Photometric uncertainty and completeness parameters.}
\label{tab:f}
\centering
\begin{tabular}{llll}
\hline \hline
Parameter & 12964(ACS) & 12964(WFC3) & 13856(WFC3) \\
\hline
$\alpha_V$ & 1.8 & 1.8 & 1.8 \\
$m_{0,V}$ & 28.1 & 28.25 & 27.70 \\
$\alpha_I$ & 1.6 & 1.8 & 2.0 \\
$m_{0,I}$ & 26.8 & 26.8 & 26.42 \\
\\
$\beta_{0,V}$ & 0.02 & 0.02 & 0.035 \\
$\beta_{1,V}$ & 0.053 & 0.056 & 0.063 \\
$\beta_{2,V}$ & 27.0 & 27.0 & 27.0 \\
$\beta_{0,I}$ & 0.02 & 0.02 & 0.025 \\
$\beta_{1,I}$ & 0.050 & 0.042 & 0.050 \\
$\beta_{2,I}$ & 25.5 & 25.5 & 25.0 \\
\hline
\end{tabular}
\end{table}


\begin{figure*}
\resizebox{\hsize}{!}{
\includegraphics{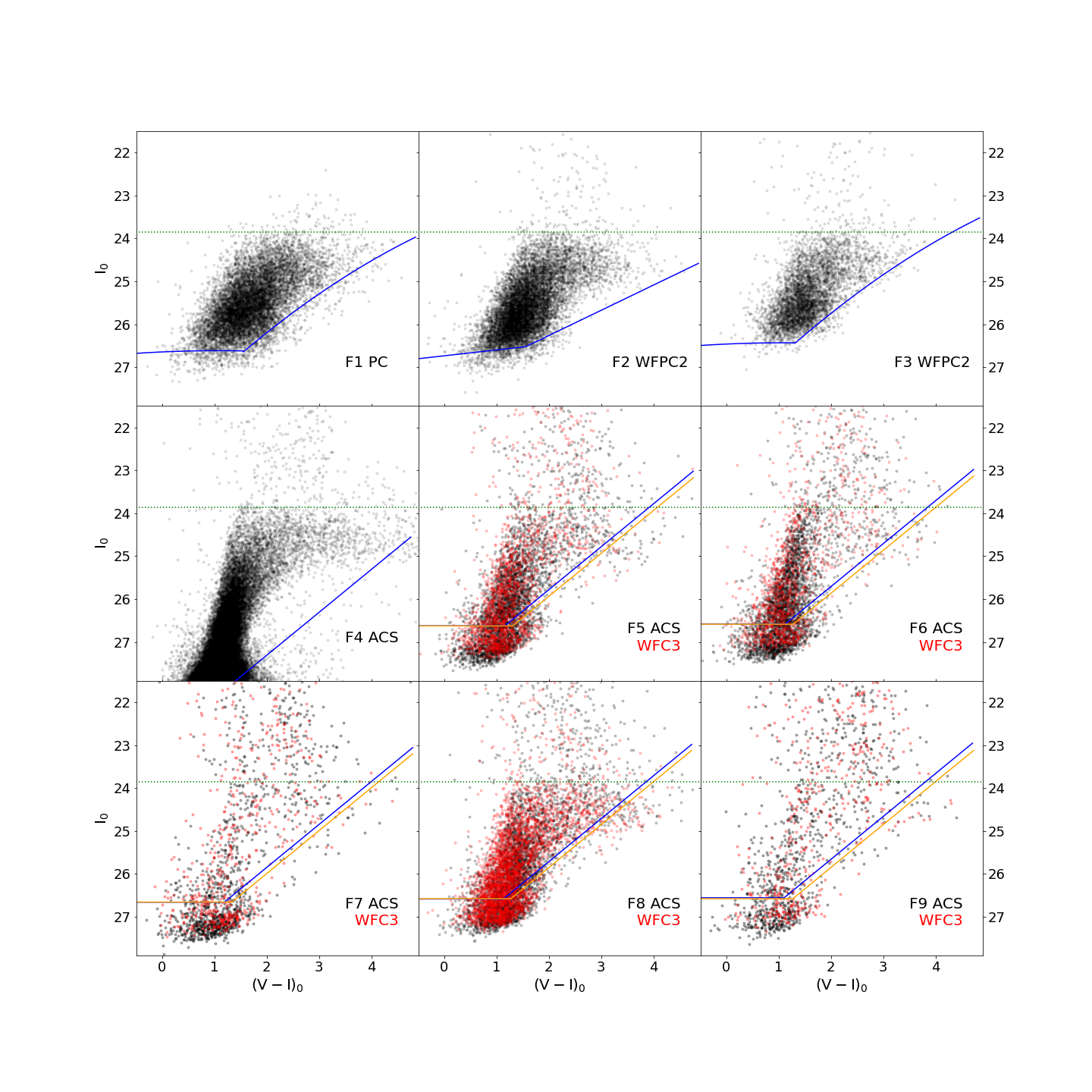}
}
\caption{Measured CMDs for fields F1-F9 as listed in 
Table \ref{tab:HSTobslog}, including the correction for Galactic extinction.  The older data for locations at 8, 20, and 30 kpc (F1, F2, and F3) were taken with
the WFPC2 camera. For the fields F5 to F9, data from ACS pointings are plotted as black dots
and and from WFC3 as red dots.  The solid lines show the 50\% detection completeness
levels (ACS and WFPC2 in blue, and WFC3 in orange). The dotted green line indicates the expected location of the TRGB. For F4, the photometric
limits are much deeper than for the other fields. }
\label{fig:cmdf1_9}
\end{figure*}

\begin{figure*}
\resizebox{\hsize}{!}{
\includegraphics{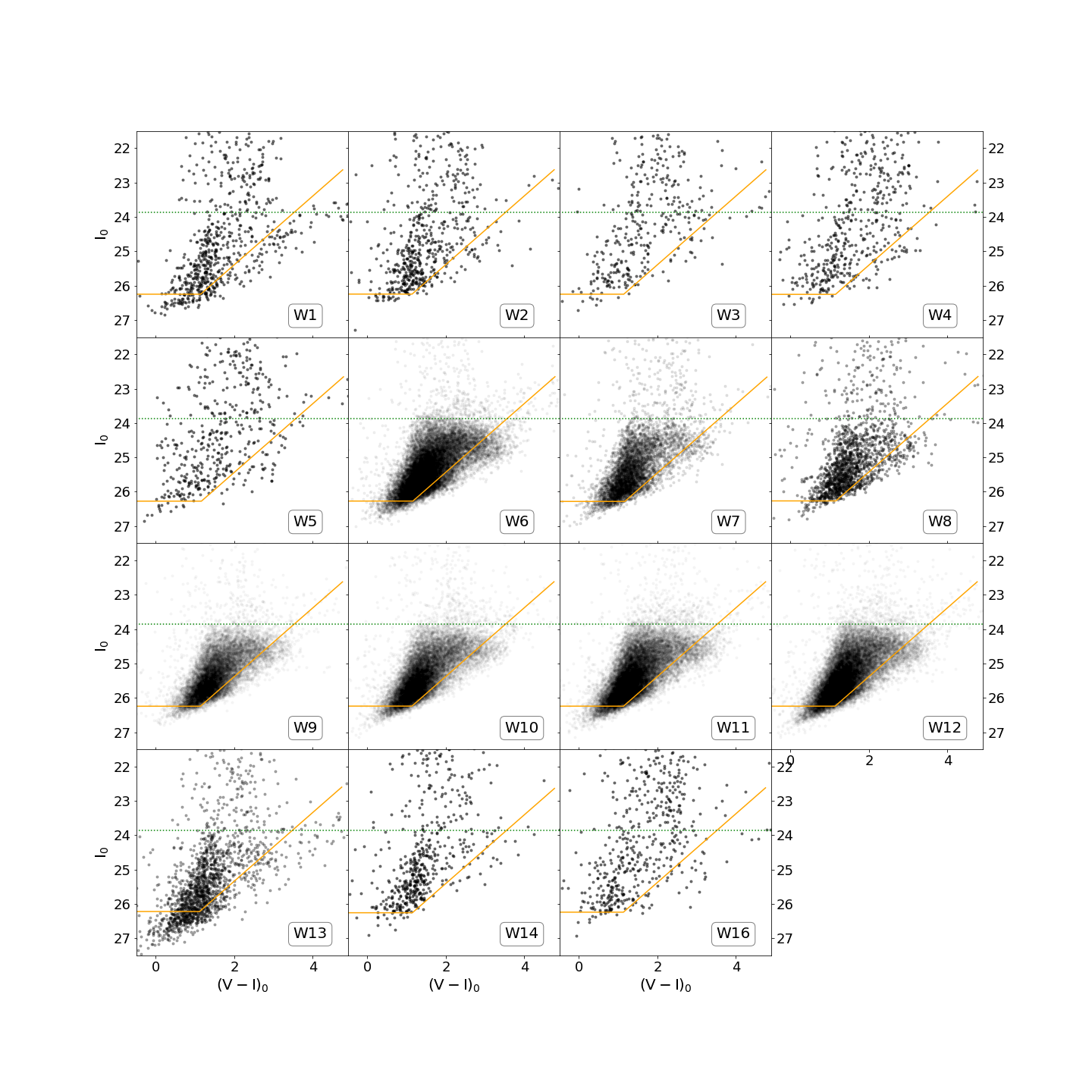}
}
\caption{CMDs for fields W1-W16 corrected for Galactic extinction.  These
fields have half the exposure time of F5-F9 and thus have brighter limiting magnitudes
and noticeably brighter completeness cutoff levels in the bluer filter. The solid (orange) lines show the 50\% detection completeness levels. The dotted green line indicates the expected location of TRGB.}
\label{fig:cmdW1_16}
\end{figure*}


The \emph{addstar} task in \emph{daophot} was used to measure the detection completeness of 
our photometry and the internal measurement uncertainties.  Artificial stars (scaled PSFs) were
added, 1000 at a time, into the images, and we then remeasured with exactly the same procedures as
described above. The completeness $f$ is the recovery fraction: the number of artificial stars detected in a given magnitude bin divided by the number of artificial stars inserted. A sample of the fraction of recovered stars versus magnitude is shown in Fig.~\ref{fig:completeness},
for the three different combinations of filter and exposure time that we have (the ACS fields from
GO12964, WFC3 from GO12964, and WFC3 from GO13856). Completeness experiments for fields F1--F4 are described in detail in the papers that reported their initial analysis \citep{harris+99, harris+harris00, harris+harris02, rejkuba+05}.

As is typical for very uncrowded images such as we have here, the detection fraction declines smoothly
from near-100\% down to near-zero over a two-magnitude run.  Analytic curves of the form
$f = 1/(1+ e^{\alpha (m-m_0)})$ \citep{harris+16} are superimposed on the data in Fig.~\ref{fig:completeness}.
In this relation, $m_0$ represents the magnitude 
where $f = 0.5$, and $\alpha$ represents the steepness of decline. 
Table \ref{tab:f} lists the pairs of parameters for the three sets of
images.  The most important feature of the adopted completeness curves 
is that they accurately describe the
shape of $f(m)$ for $f > 0.5$; we did not use any data below this
50\% point.

The exposure times and therefore also the completeness are comparable for F10A 
(GO15426) and for the Cycle 20 data from GO12964. 
As expected, with respect to these, $m_0$ is about one magnitude brighter than for the 
fields in GO13856, which have half the exposure time of the other two sets.  
As we show below, however, this difference does not seriously 
hamper our goals to derive the halo RGB population
density.

The artificial star tests were also used to determine the measurement
uncertainties of the data.  Sample artificial star runs showing the recovered magnitudes compared with
their input values are shown in Fig.~\ref{fig:delmag} for a range of magnitudes
between $\sim2$~mag brighter than the TRGB down to well below the completeness limit. These tests showed that to a close approximation,
the measurement uncertainties are about 1.4 times larger than the numbers
returned by \emph{daophot/allstar}. A useful interpolation equation for the
dependence of measurement uncertainty on magnitude is
\begin{equation}
    e_m = \beta_0 + \beta_1 e^{(m-\beta_2)} 
,\end{equation}
where the $\beta-$values appropriate for the different sets of exposures are listed in Table \ref{tab:f}.

The artificial star tests can also be used to check for the possible presence of blends. These are stars that are measured brighter due to underlying (unresolved) background and thus shift to brighter bins in the luminosity function. In the right panels of Fig.~\ref{fig:delmag} we show the difference of magnitude (measured$-$input) for stars within $\pm 0.25$~mag of the TRGB. The tail of the distribution extending to negative (brighter measured) magnitudes is negligibly small even in our most densely populated field (W12), and it is completely absent in a typical halo field (F5A). The mean magnitude of measured stars is 0.01 mag brighter for W12 field, and it shows no systematic shift for less populated fields. 
 
\section{Color-magnitude diagrams}

The CMDs for our complete set of fields are shown in Figs.~\ref{fig:cmdf1_9} and \ref{fig:cmdW1_16}. In Fig.~\ref{fig:cmdf1_9}, a small but important check on the internal consistency of our
photometry is that for each pair of ACS/WFC3 fields, the CMDs overlie each other closely, 
within the scatter of points along the RGB in each field.
Here, they match better than we found in Paper I, a direct result of the improved WFC3 color
transformations in the present work.

A broad RGB component is present in all fields, as is the clear TRGB located at $I_0 \simeq 23.86$ (indicated with the dotted green line) and
slanting downward at redder colors due to increasing bolometric correction. 
 A population of foreground stars is also clearly present in all fields over a wide range
of colors, and it extends well above the RGB population. In the outermost fields
(e.g., W3, W4, W5, and W16), the contamination appears to be comparable to the number of NGC 5128 stars.
Therefore the first necessary task for the analysis is to remove this contamination as well as
possible so that unbiased measures of the radial density distribution can be made.

The field contamination is due to a 
combination of foreground stars and very small, faint background galaxies that
managed to pass the selection criteria described above.  
Although the well-populated 
inner fields are completely dominated by the NGC 5128 halo RGB stars and thus contamination
is of relatively little concern, for our outermost fields, the objective removal of contaminants is critical.

It is possible to model the foreground population of Milky Way stars as a function of the position on the sky, but faint background galaxies provide more of a challenge. In the following, we first discuss the bright foreground stellar component in comparison with the Milky Way stellar population models. This is used to evaluate the ability of the models to reproduce the foreground Milky Way component and also to assess the possible presence of a bright AGB component in the halo of NGC~5128. A contribution from bright AGB giants is expected in some inner fields given the detection of a population of bright long period variable (LPV) stars \citep{rejkuba+03_LPVnir} and intermediate-age globular clusters (GCs) \citep{woodley+10a}, but it is so far unknown how far it extends into the halo \citep[see also the discussion in][]{crnojevic+13}. The population of contaminants in the CMD can also be evaluated empirically. In Sect.~\ref{sec:decontstat} we describe this alternative method, which we eventually preferred and adopted 
for the remaining analysis.

\subsection{Modeling the foreground component}

\begin{figure}
\resizebox{\hsize}{!}{
\includegraphics{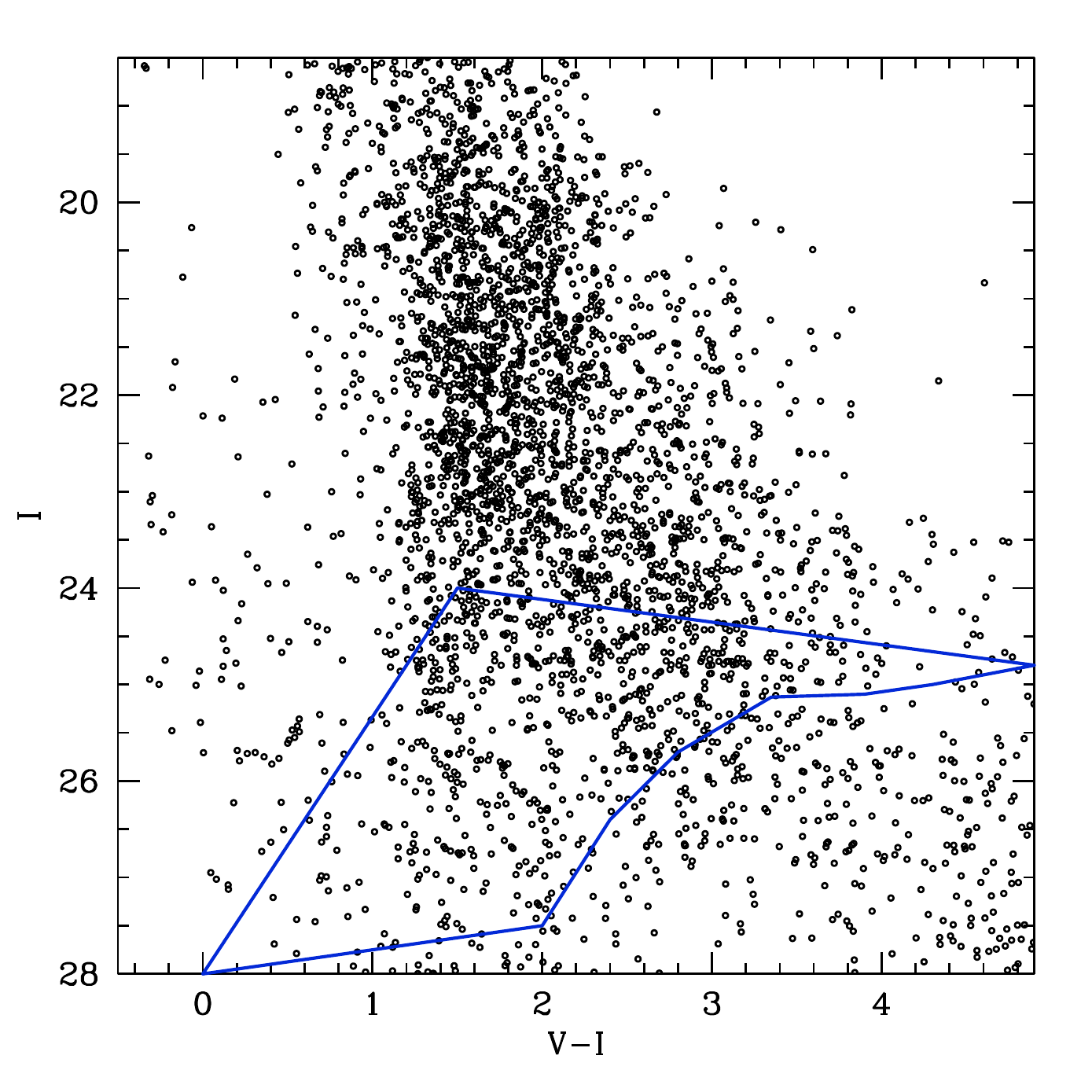}
}
\caption{CMD of the TRILEGAL simulation for a field size of 0.026 sq.deg equal to five pairs of ACS+WFC3 pointings, which best represents the foreground population (see text). The polygon appearing in Fig. \ref{fig:HST_5fieldsall}, where it encloses the majority of stars belonging to the halo of NGC~5128, is overplotted.  
}
\label{fig:cmd_fore}
\end{figure}

Models of the Milky Way stellar population can be used to simulate the number of stars and their magnitude-color distribution along the line of sight of NGC 5128. In Paper I we adopted the TRILEGAL \citep{girardi+05} simulation because it had a higher density of faint stars than the Besan\c{c}on \citep{robin+03} model. In this paper we explore in more detail the ability of the TRILEGAL model to reproduce the foreground Milky Way population in comparison with our data (see also Appendix A). In order to avoid issues with different completeness due to different exposure times, we restrict the comparison of the models to the fields observed during Cycle 20 (F5-F9).

\begin{figure}
\resizebox{\hsize}{!}{
\includegraphics{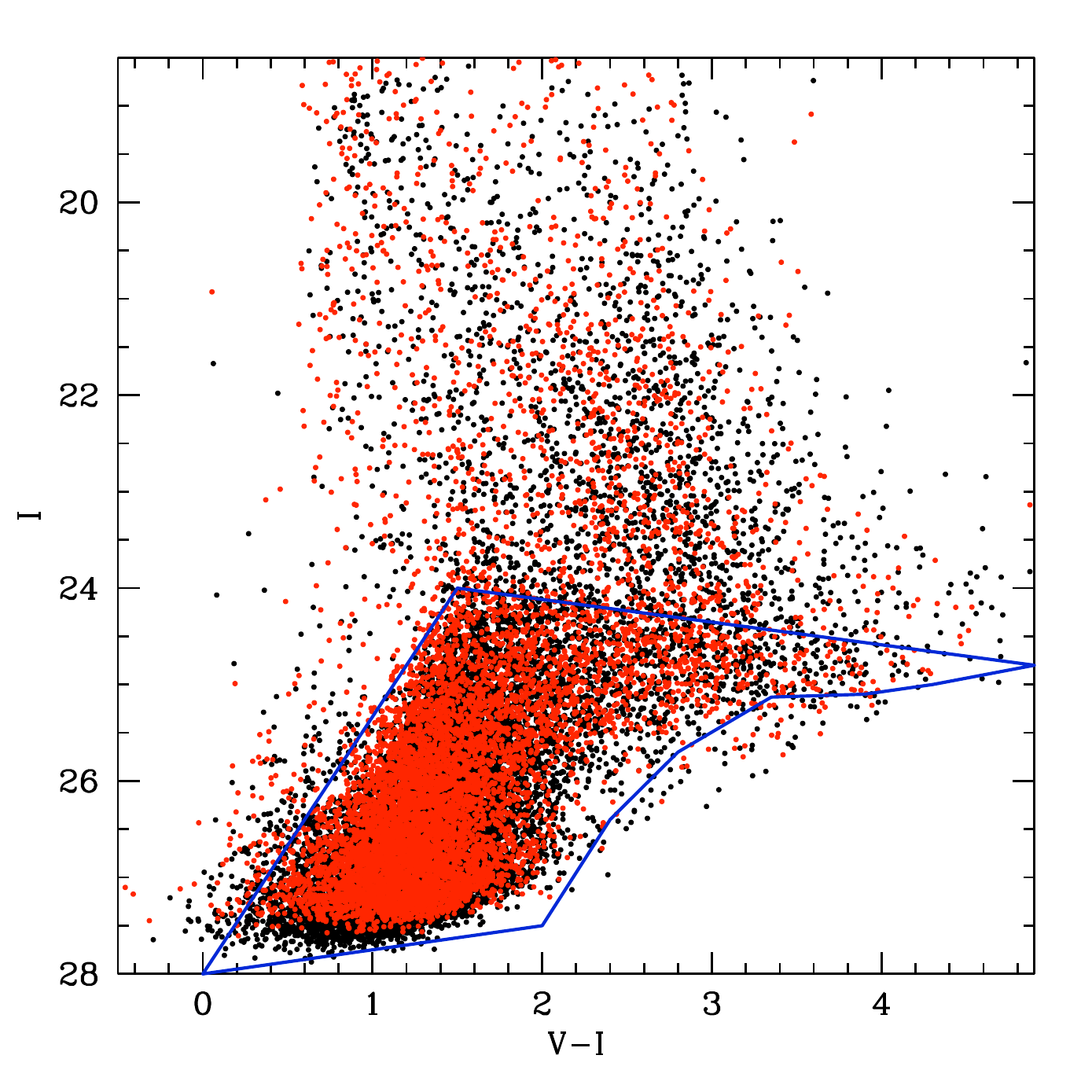}
}
\caption{Combined CMD for fields F5+F6+F7+F8+F9. Stars detected in WFC3 pointings are shown in red and in ACS pointings with black dots. The blue polygon encloses most of the stars belonging to the halo of NGC~5128.
}
\label{fig:HST_5fieldsall}
\end{figure}

The TRILEGAL simulator prompts for many different inputs. Keeping the parameters of the galactic components to their default values, we simulated a field of 0.026 square degrees, equal to the total surveyed area of 5 ACS + 5 WFC3 pointings observed during Cycle 20, in the direction of the NGC 5128 coordinates. Most of the simulated objects are main-sequence stars, but with a few white dwarfs populating the bluer colors of the distribution.

The TRILEGAL model with a Chabrier IMF and a binary fraction of 0.3 (Fig.~\ref{fig:cmd_fore}) has 1128 stars in the magnitude range $19<I<22$. In the same magnitude range, the Besan\c{c}on model has 1097 stars. The models obtained with other prescriptions have a slightly different number of stars in this range, although these differences may reflect the normalization of the TRILEGAL simulator. To compare the simulations among themselves, we normalized their counts in the range $20 < I < 22$. The simulations adopting different options (e.g., binaries on or off) yield similar results, except for the 
case of a Salpeter IMF, in which a large component of faint ($I \gtrsim 25$) and  red ($V-I \gtrsim 3$) dwarfs appears that has no counterpart on the observed CMD (Fig.~\ref{fig:HST_5fieldsall}). However, our data are largely incomplete in this faint red magnitude range in all fields except in F4, where we still do not observe large numbers of faint red dwarfs. We thus consider the Chabrier log-normal IMF and 30\% binaries as the most relevant TRILEGAL rendition of the foreground.   

To maximize the statistical significance of the foreground population for comparison with the model, we combined the CMDs for fields F5-F9, as shown in Fig.~\ref{fig:HST_5fieldsall}. In this diagram, the blue polygon encompasses the area in which the contribution of RGB stars from NGC 5128 halo dominates. 
The TRGB stands out very clearly around $I \sim 24$ mag, slanting toward fainter magnitudes at redder colors. The plume of stars brighter than the TRGB is mainly due to Milky Way contaminants having 1095 stars between $19 < I < 22$ mag. Its luminosity function (LF) is well reproduced by TRILEGAL and matches the Besan\c{c}on model very closely. In this magnitude range, the color distribution is also better reproduced by the Besan\c{c}on model. In Appendix A we examine in more detail the color distributions of TRILEGAL and Besan\c{c}on models. We find further differences with respect to observations in the bright magnitude range.

\begin{figure}
\centering
\resizebox{\hsize}{!}{
\includegraphics[angle=0,clip]{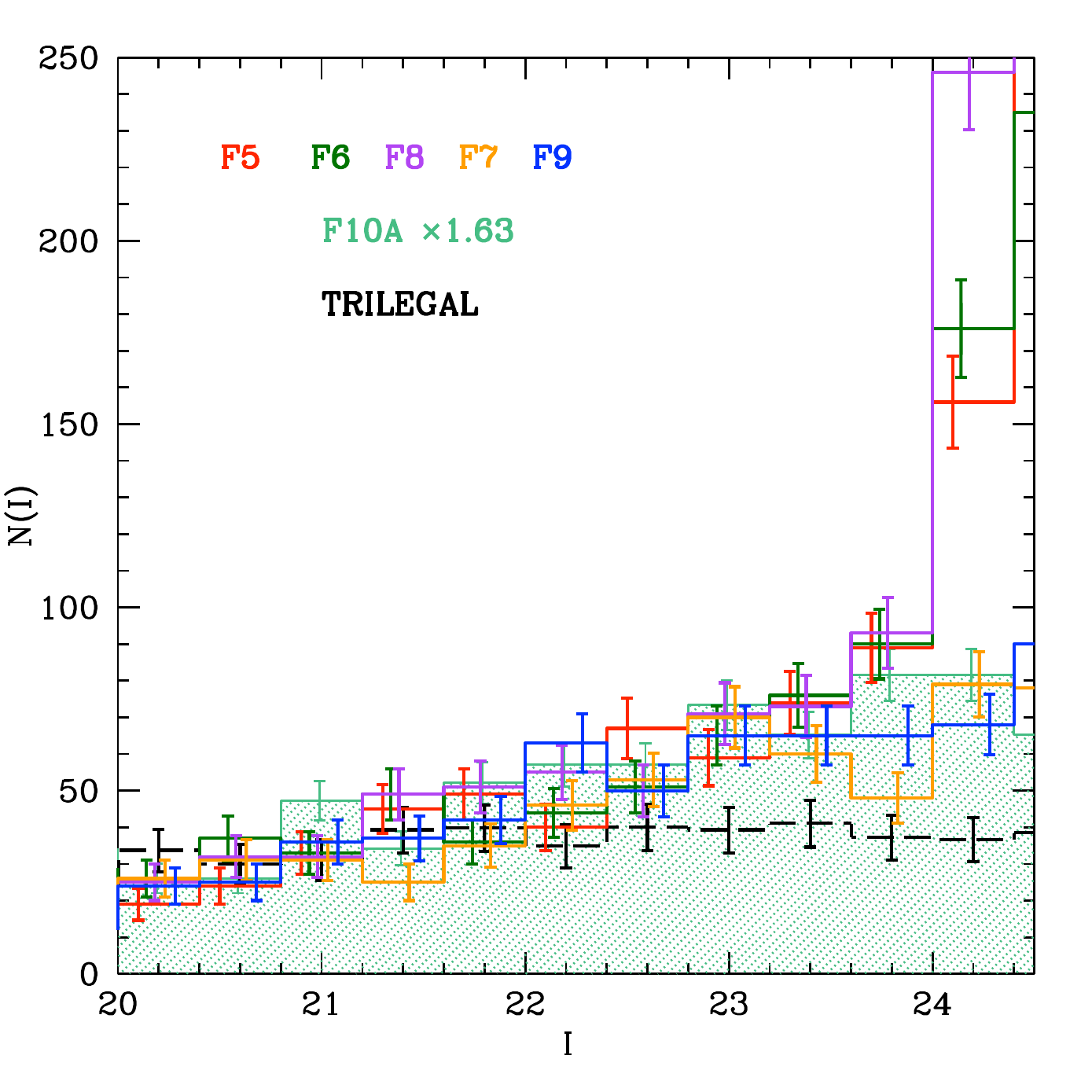}
}
\caption{Bright portion of the luminosity functions for the five Cycle 20 ACS+WFC3 field pairs, shown as the colored histograms. 
The prediction from the TRILEGAL model, normalized to the same area, is shown as the dashed black line. The dotted light green histogram shows the LF of the field F10A that best represents the empirical foreground population, appropriately scaled to the WFC3+ACS area.
}
\label{fig:lf_fields}
\end{figure}

We next investigated the LF of the stars above the TRGB.
In Fig.~\ref{fig:lf_fields} we show the observed LF data for F5 - F9, broken out for the five (ACS+WFC3) pointings.  
Each of the colored lines shows the
sum of the observed ACS+WFC3 counts, while the dashed black line shows the TRILEGAL model, scaled to the ACS+WFC3 area.
At magnitudes $I \leq 22,$ the LF is very well reproduced by the simulator.   For $I > 24,$ the observed LF increases dramatically, and we are into the main RGB population of NGC 5128. In the range $22 < I < 24,$ the data show a noticeable excess of stars with respect to the simulation, which may signal a contribution from bright intermediate-age stars in NGC 5128 or excess field contamination that is not accounted for in the model. 
On closer inspection, there is a difference between the excess counts in the three innermost 
fields (F5, F6, and F8) and in the two outer fields (F7 and F9) that sit lower. In  Fig.~\ref{fig:lf_fields} we also show as the shaded dotted light green histogram the LF of the observed field F10A (see Fig.~\ref{fig:fieldsrel} for its location), also scaled to WFC3+ACS area, which we use below for empirical removal of contamination (Sec.~\ref{sec:decontstat}).
At fainter magnitudes (I>24.5), that is, below the TRGB and not shown in this diagram, both F7 and F9 show a mild excess of stars with respect to field F10A.

In total, in these five field pairs there are 1581  objects with $22<I<24$,
while the simulator predicts 1055 foreground stars for the same area. The excess is thus
($ 526 \pm 23$) stars (where the quoted uncertainty is a lower limit due to Poisson statistics only), 
or $\sim (33 \pm 2)\%$. However, given that field F10A, which does not show any appreciable sign of RGB stars belonging to NGC 5128 in its CMD, also has an excess of star counts with respect to the TRILEGAL model, we argue that most of this excess in the outer halo fields ($R_{gc} \gtrsim 30$~kpc) is due  to field contamination that is underestimated in the model rather than to a genuine AGB component. This is further explored using the match between near-IR and optical data described in Appendix B.

The TRILEGAL simulation also predicts a component of stars fainter than the magnitude of the NGC~5128 TRGB, with colors to the red of the black polygon and fainter than the RGB. These stars, which consist of dwarfs with mass $\sim$ 0.15 M$_\odot$, consist of $\sim$ 490 objects, while on the observed CMD, there are virtually none. In this 
part of the CMD, the data are rather incomplete, which likely accounts for the discrepancy with respect to the model. Overall, the total numbers of foreground objects from the TRILEGAL simulation at magnitudes below the TRGB and within the NGC 5128 RGB polygon are much smaller than our 
sampled NGC 5128 population.  

It is important to note that in addition to the small mismatch between the observed bright star counts and the model predictions and the possible small contribution of the AGB component, the additional faint background galaxy contamination remains unaccounted for. Therefore we also explore the possibility of a strictly empirical method to subtract the field contamination in the next subsection. 

subsection{Statistical decontamination of the CMDs}
\label{sec:decontstat}

\begin{figure}
\resizebox{\hsize}{!}{
\includegraphics{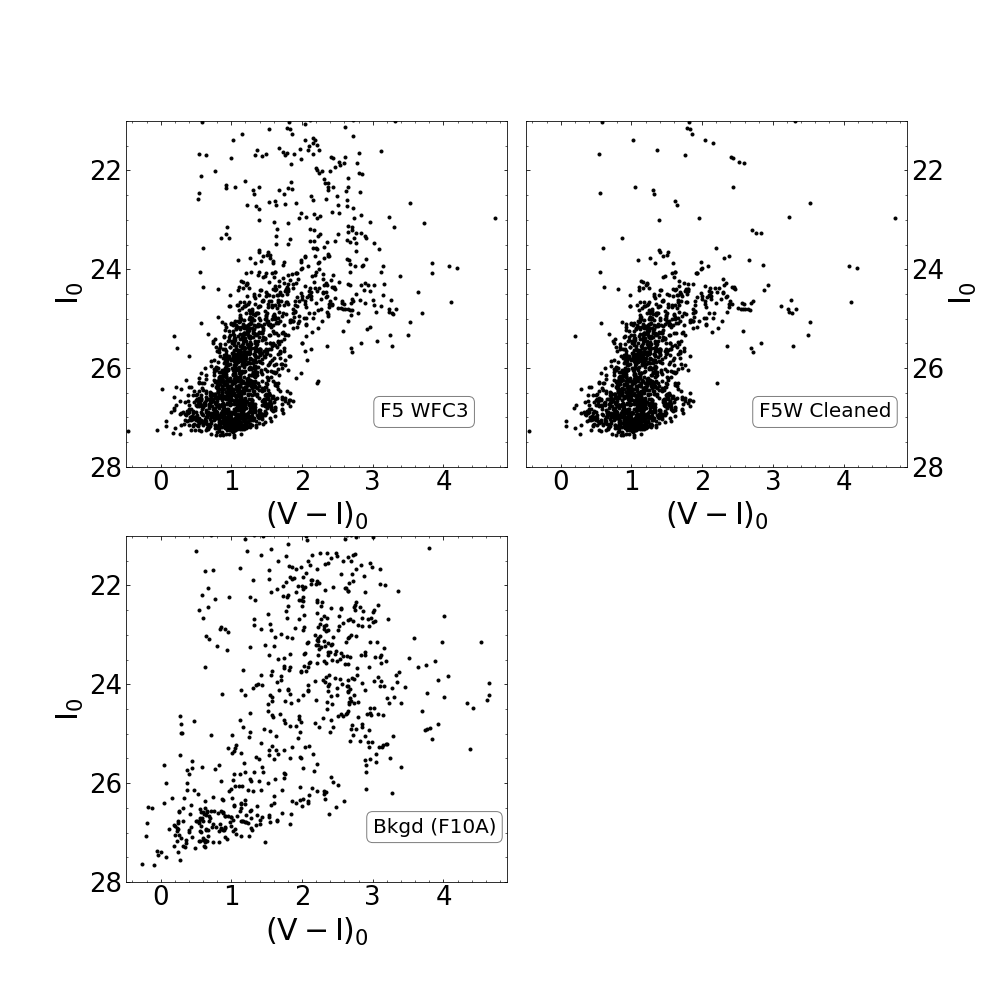}
}
\caption{Illustration of the point-by-point cleaning process described in the text.
The lower left panel shows the CMD for field F10A.  This field has little to no trace
of a clear RGB population, and we adopt it to be the control field for the remaining target fields.  In the upper panels, we show the CMD for field F5W before (left)
and after (right) cleaning. All CMDs are dereddened.  Most of the
stars above the TRGB have been removed, as have a number of other stars along and around the RGB.}
\label{fig:cleaning}
\end{figure}

\begin{figure}
\resizebox{\hsize}{!}{
\includegraphics{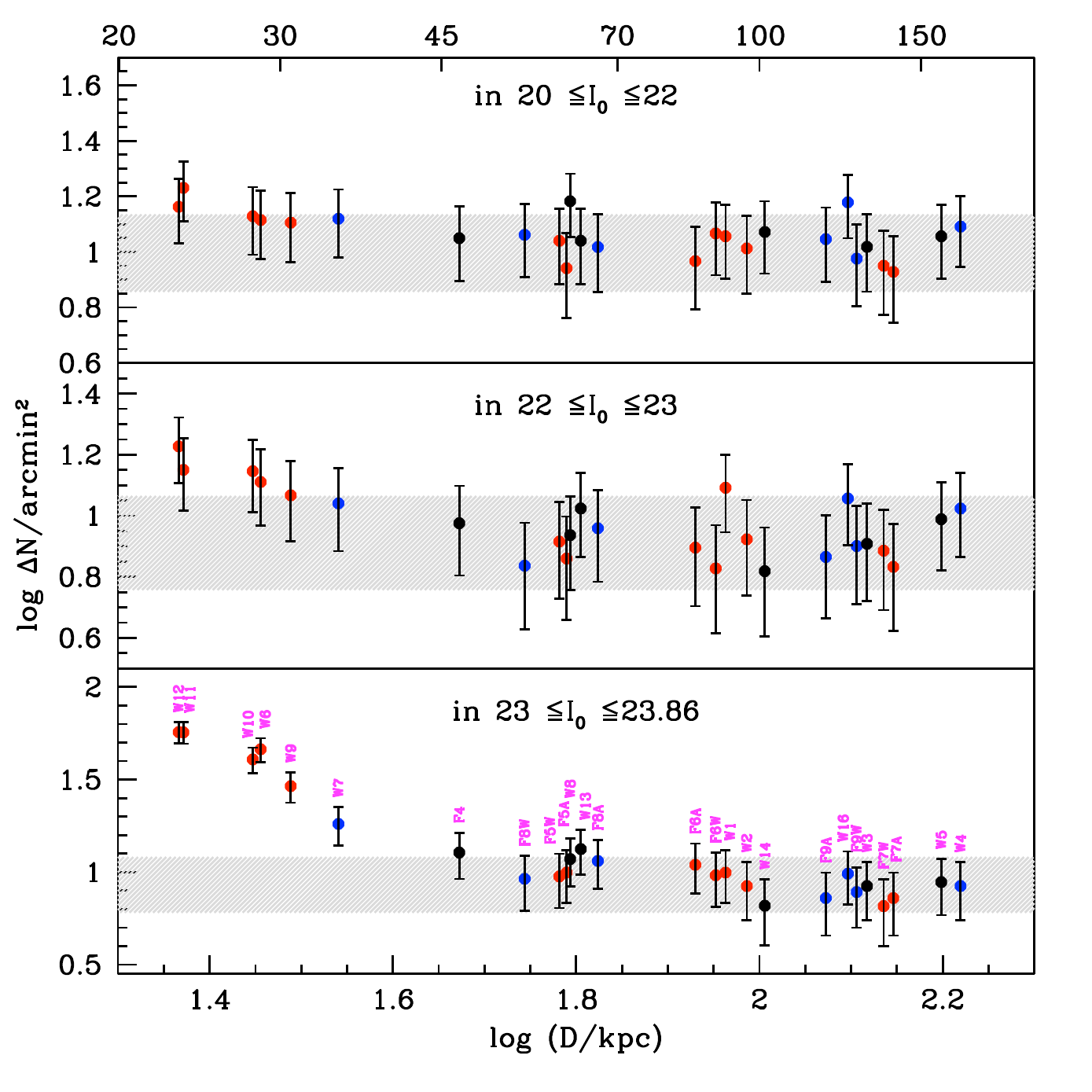}
}
\caption{Number of stars per arcmin$^2$ in all observed fields as a function of 
distance in extinction-corrected I-band magnitude bins as labeled in each panel. The bottom x-axis is the distance corresponding to the semimajor axis of the elliptical isophote that goes through the field (assuming ellipticity of the outer halo e=0.54 for all fields) in log and top x-axis in linear scale. Red (blue) dots highlight fields along, or close to, the major (minor) axis, and fields plotted with black dots are in between. The hashed horizontal band in each panel indicates the number of stars (with 1$\sigma$ Poissonian error) in the background field F10A in the same magnitude bin.}  
\label{fig:counts_bright}
\end{figure}

Ideally, the preferred and purely empirical
way to determine the field contamination independently of any modeling 
would be to use HST pointings just outside the NGC 5128 halo, observed with the same filters and similar exposure times to those used in our program.\footnote{In our original observing program
design, we anticipated that our most remote targets, such as F7, would give us 
suitable control fields for this purpose.  However, the halo proved to be much more
extended than expected (at least along the NE axis).}

An extensive search of the MAST Archive for fields matching these criteria was unsuccessful,
\emph{} except for field F10A (see Table \ref{tab:HSTobslog} and Fig.~\ref{fig:fieldsrel}).  At a projected distance of $R \sim 120$ kpc from the
center of NGC 5128, it is on the opposite
side of the galaxy (to the southwest) from our other outer fields.  The CMD for F10A is shown
in the lower left panel of Fig.~\ref{fig:cleaning}.  It reveals a swath of stars crossing the
diagram in the upper right corner. These stars are mostly foreground Milky Way stars, which appear in every one of our other CMDs, and they are also clearly visible in the TRILEGAL model of the Milky Way (Fig.~\ref{fig:cmd_fore}).  The F10A CMD also contains some very faint objects in the lower left corner, but little else.
There is little or no clear trace of an RGB population that would be found in the
range $I = 24-26$ and $(V-I) \sim 1 - 2$, as is the case in the CMDs for
all other fields.  Of all the HST fields in the region that are available to us,
it has the lowest population of stars in the RGB region that is our target.  The fact that it is
not quite as distant from the galaxy center as our fields F7 and W5 and yet has a smaller
population of stars suggests to us simply that
the halo of NGC 5128 is not ideally smooth and symmetric at these outermost distances. 
More or less of necessity, we adopted F10A as our control field.

Because our target fields are spread over a large area spanning 
almost 4 degrees in diameter (Fig.~\ref{fig:fieldsrel}), differences in the
number density of contaminants larger than simple Poisson statistics may be present. 
Neither the foreground stars nor the faint background galaxies are
uniformly spread across an area this wide.
That is, the assumption of a single mean contamination level is
an approximation.  This field-to-field variance sets an ultimate limit on our
ability to decontaminate the CMDs using a single background field.

Figure~\ref{fig:counts_bright} shows the number density of bright stars in all the fields with respect to our nominal background field F10A (see below). Stars are selected according to the specified magnitude bins without further selection by color, but the correction for incompleteness is applied according to parameters given in Table~\ref{tab:f}. Between $20<I_0<23$, that is, more than one magnitude brighter than the TRGB, the field-to-field variations in the number of Milky Way foreground contaminants are fairly consistent within $\sim 1-2 \sigma$. This gives us confidence that statistical decontamination with a single F10A field works. There is also a notable excess in surface density in the bottom, and more marginally, also in the middle panels in Fig.~\ref{fig:counts_bright} in the inner halo fields. We return to this in Sect.~\ref{sec:AGB}.


\begin{figure*}
\resizebox{\hsize}{!}{
\includegraphics{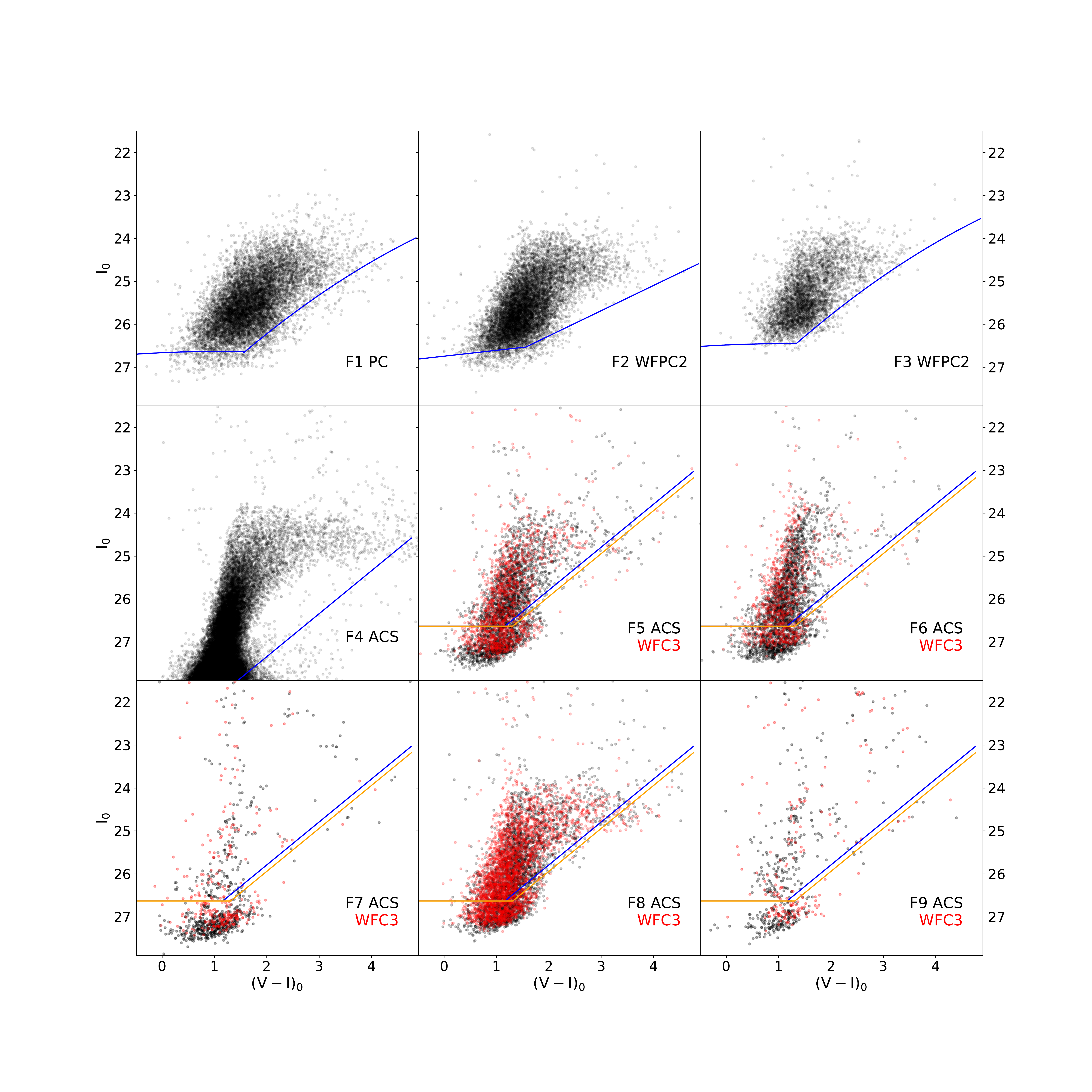}
}
\caption{CMDs for fields F1-F9 after the point-by-point fore- and background decontamination described in the text. Data from ACS pointings are plotted as black dots,
and those for WFC3 as red dots.  The solid lines show the 50\% detection completeness
levels (ACS and WFPC2 in blue, WFC3 in orange).}
\label{fig:cmdf1_9_clean}
\end{figure*}

\begin{figure*}
\resizebox{\hsize}{!}{
\includegraphics{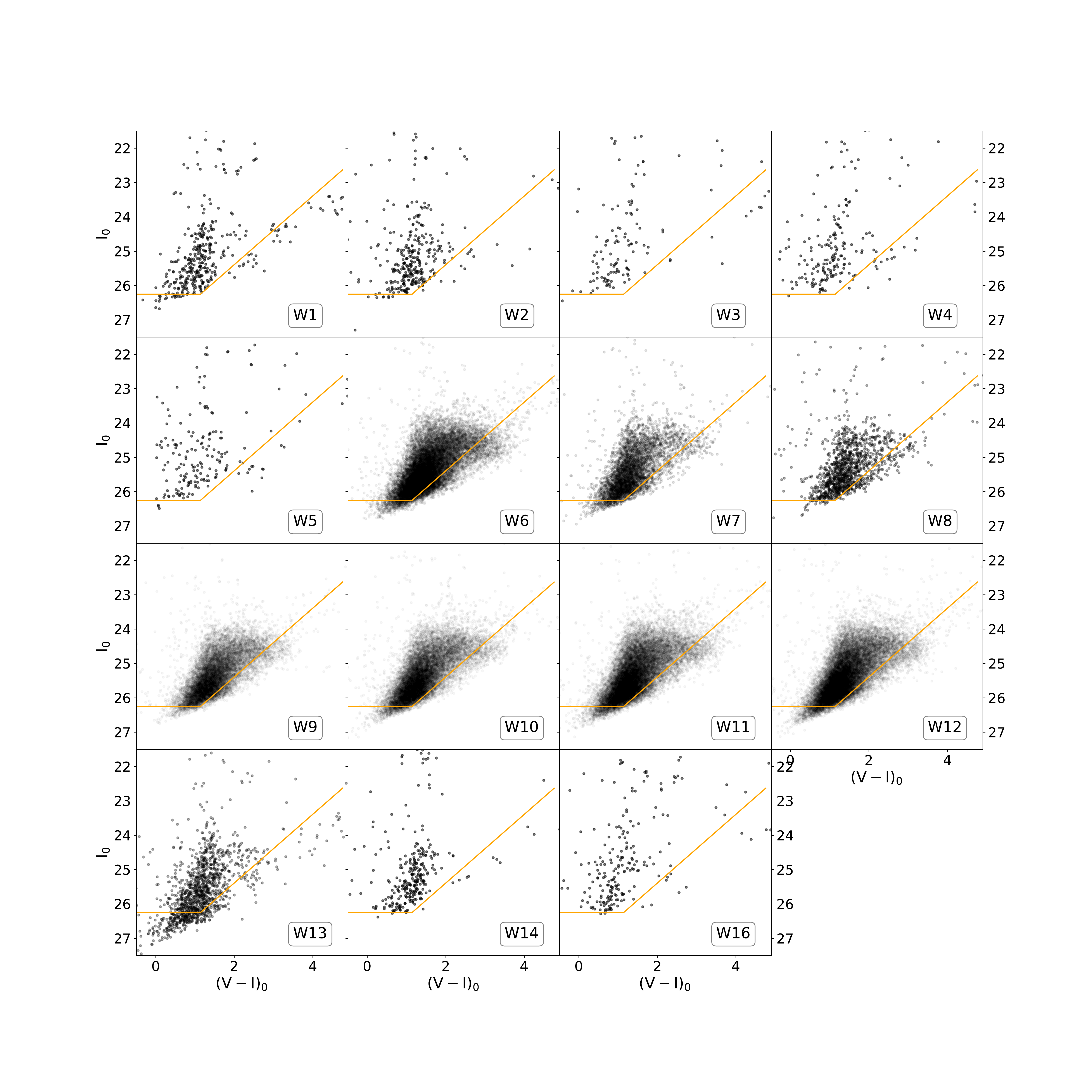}
}
\caption{CMDs for fields W1-W16 after cleaning. The solid (orange) lines show the 50\% detection completeness levels.}  
\label{fig:cmdW1_16_clean}
\end{figure*}


With the control field F10A in hand, star-by-star subtraction of its starlist
was performed in the conventional way by searching for nearest
neighbors in the two CMDs and matching them.  Specifically, we assumed that the target field 
(the one needing cleaning) consists of a list of stars with magnitudes and colors ($m_i, c_i$).  Similarly,
the background field consists of a list ($m_j, c_j$).  Each star 
in the target field list is surrounded by an ellipse of axial
length $x_c$ in $(V-I)$ and $x_m$ in $I$.  Its nearest
neighbor in the control field list is then found, where the distance $d$ between
them is defined as
\begin{equation}
d^2 =  \frac{\Delta(V-I)^2}{x_c^2} + \frac{\Delta I^2}{x_m^2}.
\end{equation}
\noindent If that neighbor
falls within the ellipse ($d_{min} < 1$), then both stars are removed from their lists.
After experimentation with different ellipse axes, we adopted
$x_c = x_m = 0.2$ mag.  A key consistency test of the method is that
the stars in the CMD brighter than the TRGB ($I < 24$, i.e., the
region of the CMD that is almost entirely contaminants) should be 
successfully removed to within Poisson $\sqrt{n}$ statistics, where $n$ is the
number of control-field stars in that part of the CMD.

F10A is an ACS field, so its entire starlist was used for cleaning the other
ACS fields F4-F9.  For the WFC3 fields, which have 0.64 times the area of ACS,
64\% of the stars in F10A were selected at random.
Before the CMD matching was made, the control field and target field starlists 
were both individually dereddened according to $E_{V-I} = 1.24 E_{B-V}$ and
$A_I = 1.55 E_{B-V}$ and with the different
foreground reddenings of the fields, as given in Table \ref{tab:HSTobslog}.
The method is illustrated in Fig.~\ref{fig:cleaning}.
The results are shown for field F5W, which has a clear RGB component, but
also a noticeable contamination fraction.  

The CMDs for all the program fields after this point-by-point cleaning
are shown in Figs. \ref{fig:cmdf1_9_clean} and \ref{fig:cmdW1_16_clean}.
As suggested above, the most noticeable changes
in the CMDs are in the bright regions ($I < 24$), but the RGB populations
in the sparsest fields (such as F7, F9, W3, W4, W5, W16) now also stand
out more clearly.  
It is also clear that some residual contamination is still present;
in relative terms, this is most noticeable in the sparsest fields.  
For these sparse outermost fields, a very small amount of ``noise'' (the contaminants) must be subtracted from a similarly small
``signal'' (the RGB stars), so the results are strongly subject to small-number
statistical fluctuations.  In addition, as noted above, both components show physical density
differences across the entire halo that do not strictly
follows Poisson $\sqrt{n}$ statistics.  In short,  
the cleaning process for this problem was limited by the intrinsic field-to-field statistical
variance of the halo stellar populations and the background and not by the details of our cleaning method.

To gain an alternative look at a mean pure-background population of stars, we also experimented with
an iterative approach using the target fields themselves.  This approach takes 
the starting point that the number density $\phi$ of stars in each field has
two components:  the RGB stars belonging to NGC 5128, and a contaminant
population,
\begin{equation}
    \phi = \phi_{RGB} + \phi_{bkgd}  .
\end{equation}
\noindent We then assumed (necessarily, although not quite correctly) that $\phi_{bkgd}$ is the same 
for all the fields, and also that $\phi_{RGB}$ follows a power-law
decline with radius $\phi_{RGB} \simeq R^{-\alpha}$ , as found in Paper I.
Assuming that the power-law slope $\alpha$ remains known and constant at all $R$
then gives enough leverage to solve for $\phi_{bkgd}$.  In practice, we took
one of the sparsest outer fields (such as F7 or F10A) and subtracted its CMD from one
of the inner fields that was more dominated by the RGB component (such as F5). This left
a CMD that contained only RGB stars to within the field-to-field statistical scatter.  This 
pure-RGB CMD was then scaled by $(R, \alpha$) to subtract the RGB
component from any other field, leaving only $\phi_{bkgd}$.  Iteration of this approach yields much
cleaner and contamination-free CMDs.  However, the assumption of
a single power-law form for $\phi_{RGB}$ imposes a rather strong condition
on the approach, and it did not yield results that were different from
or better than the purely empirical approach with the single control field
described above.

For the purposes of the present paper, we adopted the cleaned CMDs as determined
from the use of only F10A as a control field. Our conclusion is that we can trace
the RGB population outward to the point where $\phi_{RGB} \sim \phi_{bkgd}$,
as can be seen by comparing, for example, the cleaned versions of F7, F9, W3, or W5
in Figs.~11-12 with their original CMDs in Figs.~4-5.
Tracing the halo to still larger radii requires star counts that reach deeper magnitude levels than we have at present, and that cover a wider area.


\begin{figure*}
\centering
\resizebox{\hsize}{!}{
\includegraphics[angle=0,clip]{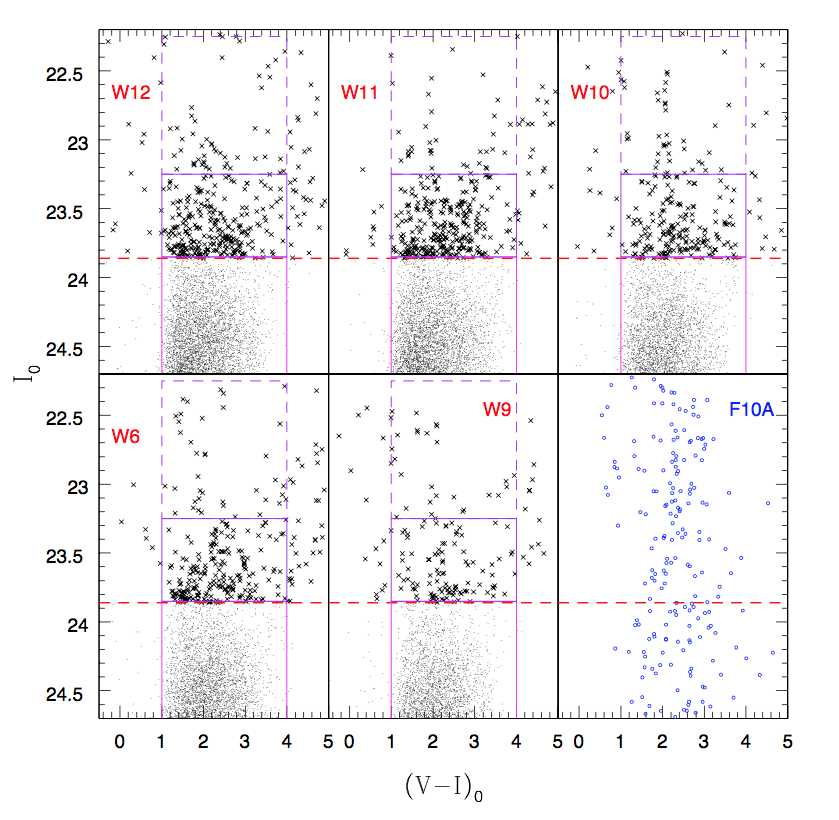}
\includegraphics[angle=0,clip]{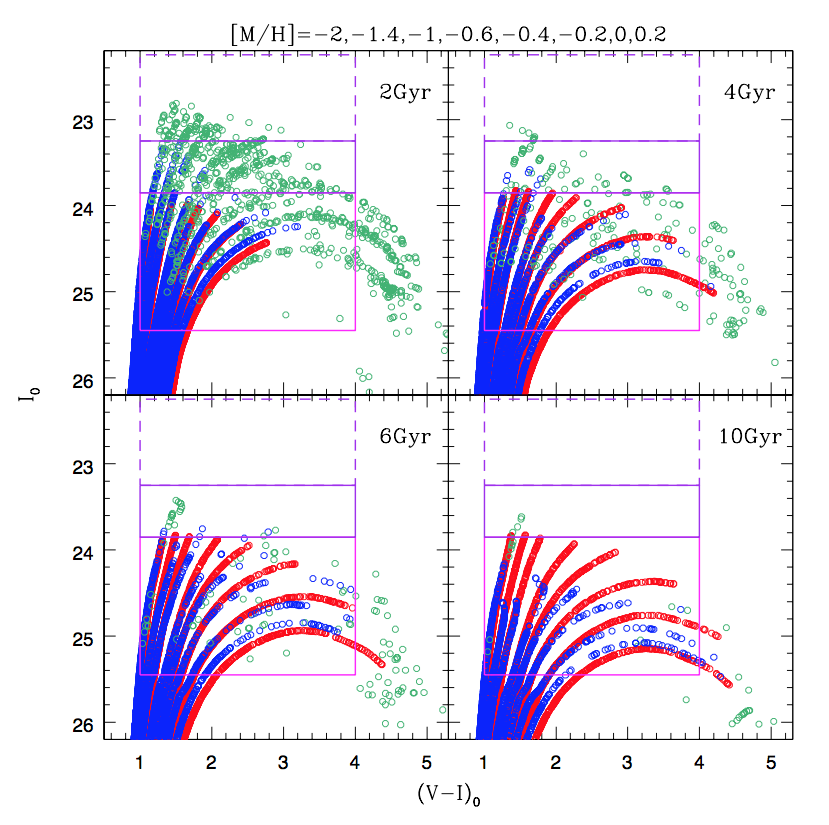}
}
\caption{Bright portion of CMDs of the five fields that show the highest excess of stars above the TRGB with stars brighter than the TRGB (dashed horizontal line at $I_0=23.86$) plotted as crosses (left). These CMDs have been statistically cleaned from foreground contamination using observations in F10A shown in the lower right CMD. Padova stellar evolutionary isochrones for different metallicities listed at the top, and ages 2, 4, 6 and 10 Gyr (right). The isochrones have been shifted to the distance of NGC~5128 and are color-coded according to stellar evolutionary stage: RGB (red circles), early AGB (blue circles), and TP-AGB (green circles). The boxes indicate regions for which the mass-specific production factor $P_j$ has been computed from the isochrones (see text).}
\label{fig:cmd_isochs}
\end{figure*}

\section{Intermediate-age AGB component in the inner halo}
\label{sec:AGB}

\begin{figure}
\resizebox{\hsize}{!}{
\includegraphics{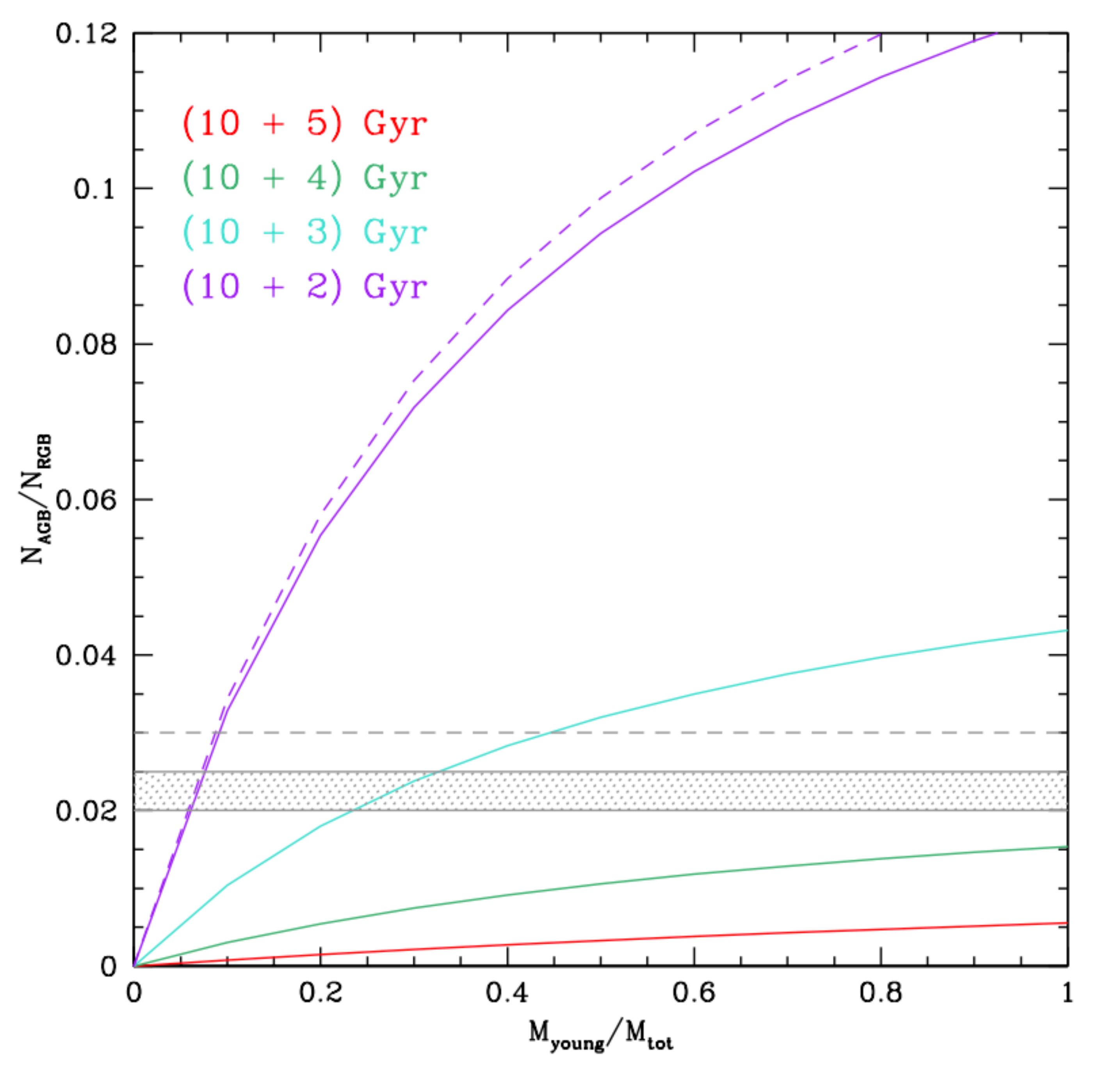}
}
\caption{Number ratio of the bright AGB vs. RGB stars selected within the boxes described in Fig.~\ref{fig:cmd_isochs} for a theoretical stellar population including two bursts (see legend) as a function of mass of the younger population. The shaded band encompasses the ratio of stars in our inner halo fields (W6, W9, W10, W11, and W12) counted in the boxes plotted with solid lines in Fig.~\ref{fig:cmd_isochs}. The horizontal line is the observed ratio when brighter stars are also included within the dashed box ($22.25<I_0<23.85$) with respect to counts in the RGB box ($23.85<I_0<25.45$).}
\label{fig:plratio}
\end{figure}

In the bottom panel of Fig.~\ref{fig:counts_bright}, within the magnitude bin $23\leq I_0 \leq 23.86$ that is just above the tip of the RGB, the increase in the observed surface density of stars in the fields located within $\sim 30$~kpc from the center is strong. The excess is strongest in fields W11 and W12, which are nearly collocated at about 23 kpc galactocentric distance along the southwestern major axis. The nearby fields W9 and W10, located at $R_{gc}\sim$29 and $\sim$28~kpc along the southwestern major axis, respectively, and field W6, which is at a similar distance as W10 ($\sim$28 kpc), but on the other side of the galaxy along the northeastern major axis (see~Fig.~\ref{fig:fieldsrel}), show a slightly lower, but still highly significant excess of stars above the TRGB. Counts in the same magnitude bin in W8 and F4, which are at $\sim 37$~kpc in between major and minor axis toward the southeast and south, have stellar densities at the upper limit of the foreground field F10A, but are consistent with its stellar density within $1\sigma$. The innermost field, W7 at $R_{gc}\sim 20$~kpc close to the minor axis, also shows an excess of stars brighter than the TRGB, although only at the $\sim 2\sigma$ level. Considering projected distances on sky alone (assuming spherical halo), it appeared that the bright star excess is lowest in W7, the field that has lowest projected galactocentric distance. However, after accounting for a higher ellipticity of the halo of e=0.54 (see Sec.~5) and arranging fields according to their distances that correspond to the semimajor axis of the elliptical isophote that goes through the field, the minor axis field W7 is at 34.7~kpc (see Fig.~\ref{fig:counts_bright}). This shows that when the ellipticity in the halo is accounted for, the bright star excess increases monotonically inward. Because of older cameras used for observations in F1-F3 and their different completeness and crowding corrections,
we decided to exclude these pointings from this analysis.\footnote{F3 located at distance corresponding to $\sim 38$~kpc, when accounting for ellipticity, shows bright star excess intermediate between W7 and F4, thus continuing the observed trend. F2, located near the same isophote as W6 and W10, has significantly fewer stars than these two fields. F1, the innermost field at 1.8~Re, has a very high excess of bright stars, but part of this might be due to blends. }

This excess of stars above the TRGB might be due to bright AGB giants in NGC~5128 or to a contribution from an additional foreground component. It might also be due to scatter of stars from the top of the RGB due to photometric errors. We examine each option in turn. 

\subsection{Blends and photometric errors?}

Larger photometric errors and scattering of stars into a brighter magnitude bin due to blends of two or more RGB stars close to the TRGB would result in an excess of stars just above the TRGB, as observed in the bottom panel of Fig.~\ref{fig:counts_bright}. We note, however, that fields  showing this excess also have a marginal excess in the brighter magnitude bin ($22<I_0<23$, middle panel), which cannot be populated in the same way because too few stars lie around $I_0 \sim 23$ and the photometric errors are smaller. Furthermore, our fields are not crowded, and our extensive completeness tests exclude the possibility of such large errors and blends (see Fig.~\ref{fig:delmag}).
Therefore this hypothesis is rejected.

\subsection{Additional foreground component?}

An additional foreground component may seem a more plausible explanation for the excess of contaminants in W9+W10+W11+W12, which are very close on the sky. We combined the $I_0$ LF of these four nearby fields and used the Sobel filter to measure the discontinuity in the LF, which could correspond to the TRGB. While the TRGB belonging to the NGC~5128 is clearly detected at $I_0 = 23.85 \pm 0.02$~mag, the brighter stars show a quite sharp increase in the LF at $I_0= 23.3 \pm 0.1$~mag. If this sharp feature in the LF were due to a TRGB, this would place the foreground object at a distance of 2.95~Mpc, compatible with a dwarf satellite in the Centaurus A group \citep{mueller+19}. Indeed, NGC~5128 is known to host a rich system of dwarf satellites \citep{karachentsev+07, mueller+17,crnojevic+19}. However, at the distance of 2.95 Mpc W11-W9 are separated by $4.88$~kpc, which is much too large for a dwarf galaxy. The similar excess observed in W6 is even more problematic. W6 is even farther away on the other side of the galaxy. The physical distance between these fields makes the hypothesis of a foreground object implausible. 
We also checked the dependence of the excess of stars on the MW latitude and found no correlation. Based on these arguments, we deem it unlikely that foreground contamination primarily causes the observed bright stars in the inner halo fields.

\subsection{Intermediate-age AGB component}    
Intermediate-age AGB stars could explain the excess counts above the TRGB. We note that in addition to the excess in stellar density in the bin immediately above the TRGB, as seen in the bottom panel in Fig.~\ref{fig:counts_bright}, the middle panel also indicates a mild excess in the next brighter bin  ($22 \leq I_0 \leq 23$) in the same fields. 

Figure~\ref{fig:cmd_isochs} compares the bright portion of the statistically decontaminated CMDs in the fields that show excess counts (left panels) with the theoretical isochrones, which are shifted to the distance of NGC~5128. The isochrones were constructed using the CMD tool\footnote{\tt{http://stev.oapd.inaf.it/cgi-bin/cmd}} maintained by L. Girardi and are based on the Padova set of stellar evolutionary models that match PARSEC isochrones \citep{bressan+12} ending at the first thermal pulse with COLIBRI models \citep{marigo+13} that describe the thermally pulsing AGB (TP AGB) phase. The latter are calibrated on the characteristics of the population of AGB stars in the Magellanic Clouds 
\citep{pastorelli+19, pastorelli+20}. The theoretical isochrones are plotted for the models ranging from the most metal-poor ([M/H]$=-2$~dex; bluest model) to super-solar metallicity ([M/H]$=+0.2$~dex; reddest model). The red dots show RGB, blue show early AGB, and green presents the TP AGB evolutionary phase. The 
density of points 
along the isochrones is proportional to stellar evolutionary lifetime: Each isochrone is a simulation of a simple stellar population with a given age (2, 4, 6, or 10 Gyr) and metallicity, as indicated in the panels. The boxes drawn over the isochrones are the same as those overplotted on the observed CMDs on the left. The magenta box (hereafter RGB box) spans the magnitude range of the upper RGB up to the TRGB ($23.85<I_0<25.45$), the solid purple box (hereafter AGB box) samples 0.6 mag above the TRGB ($23.25<I_0<23.85$), and the dashed purple box spans a brighter range $22.25<I_0<23.85$. The isochrones show that for old ages, only the most metal-poor populations may contribute stars brighter than the TRGB. There are in any case too few of these and they are too faint to account for the excess observed in the inner fields. Clearly, a younger population of 4 Gyr or younger is needed to account for the observed excess of stars. 
 
We can estimate the contribution of this young component using the observed star counts ($N_j$) in the RGB and AGB boxes and the theoretical mass specific production in the same boxes ($P_j = N_j/M_{SSP}$; \citealt{greggio+renzini2011}). The $P_j$ factors, derived from the simulated simple stellar populations, multiplied by the number of observed stars in the diagnostic boxes, yield the initial mass of their parent SSP \citep{greggio02}.\footnote{The model populations were computed assuming a Salpeter IMF, but the IMF is irrelevant for the computation of the $N_{AGB}/N_{RGB}$ as long as it is the same as for both components.} 

We find that younger stellar populations are more efficient producers of stars in the RGB box, that is, they provide more objects per unit mass in this location on the CMD. The dependence of $P_{RGB}$ on the metallicity is instead weak. Conversely, $P_{AGB}$ shows a strong dependence on metallicity at ages younger than $\sim$ 4 Gyr, with a peak at $[M/H] \simeq -0.5$.
Adopting a median metallicity for the stars in the inner halo of NGC~5128 of [M/H]$=-0.5$~dex (see Paper~I), we computed the expected ratio of star counts in the AGB and RGB boxes as function of the contribution of the young component to the total mass of the stellar population sampled in the field for different combinations of the ages of the old and the young component. Figure~\ref{fig:plratio} shows the results for a 10 Gyr old stellar population coupled with a young component of various ages. The shaded area represents the range of this ratio as measured in the inner fields under discussion. Dashed lines refer to the case in which the AGB box extends up to $I_0 = 22.25$.
The observed $N_{AGB}/N_{RGB}$ ratio is compatible with a ~10\% fraction of $\sim 2$~Gyr component along with an older population. Alternatively, an even larger ($\sim 30-40$\%) fraction of 3 Gyr old stars is required. It is impossible to reproduce the observed counts above the TRGB without a population of at least 3 Gyr or younger.

Based on the above, we conclude that in addition to the bulk of the old stellar population, there is the excess of stars above the TRGB within ($\la 35$~kpc) due to an intermediate-age AGB component with 2-3 Gyr old stars contributing between 10-30\% in mass. In the fields within $\la 25$~kpc, which show excess also in the $22 < I_0 < 23$~mag bin, the younger population may contribute even up to 40\% of mass if its age is 3~Gyr. 

This result is a fourth piece of evidence that the inner halo of NGC~5128 had an extended star formation or multiple star formation bursts. The variability period analysis of the luminous 
LPV stars discovered in two fields located $\sim 18$~kpc to the northeast and $\sim 10$~kpc to the south \citep{rejkuba+03_LPVcat} revealed a fraction of $\sim 10$\% of LPVs with periods longer than 500~days, which are due to more massive and hence younger progenitors with ages ranging between 1.5-5 Gyr, depending on the  metallicity \citep{rejkuba+03_LPVnir}. The spectroscopic study of 
GCs using Lick indices \citep{woodley+10a} showed that 68\% of the NGC 5128 GCs have old ages ($>8$ Gyr), 14\% have intermediate ages (5--8 Gyr), and 18\% are younger than $5$~Gyr. Finally, \citet{rejkuba+11} compared the deep CMD of F4, which reaches the core helium burning red clump stars, with simulated CMDs.  They found that two-burst models with 70--80\% of the stars formed $12\pm 1$ Gyr ago and 20--30\% contribution of $2 - 4$ Gyr old stars provide the best agreement with the data. Almost all GCs are located within $\sim 30$~kpc, simply because this was the area of the galaxy explored at the time. More recent surveys \citep{taylor+17_SCABS, hughes+21} have found many more GC candidates extending out to more than 100~kpc, but not much is known about their age and metallicity distributions.
We note that no high excess of bright AGB stars is present in F4; see the bottom panel of Fig.~\ref{fig:counts_bright}, which shows that this field has a stellar density just above the upper range of our foreground field F10A. The two results agree if the majority of the intermediate-age stars in this field are between 3-4 Gyr old because we showed that stellar populations older than 3 Gyr do not contribute much to the bright AGB counts (Fig.~\ref{fig:cmd_isochs}), except for the very metal-poor ones, which are not strongly present in the field F4 \citep{rejkuba+05}.  

Given the active assembly history of NGC~5128, it is not surprising to find such an intermediate-age population fraction  even out to $\sim 30$~kpc in the halo. In a recent paper, \citet{wang+20} proposed that NGC~5128 is a result of a major merger, based on simulations that started 6 Gyr ago with two nearly equal-size progenitors and ended with a fusion 2 Gyr ago. The simulations reproduced several observables fairly well (e.g., planetary nebula velocity field, morphology of the halo and that of shells and streams observed in the PISCeS survey, metallicity distribution, halo metallicity gradient, and star formation in the central disk of NGC~5128), and made predictions that do not have sufficient observables to constrain the model. One of these is the stellar age distribution in the halo, which shows an increase of intermediate-age stars and younger mean age in the inner halo, in qualitative agreement with our finding (for more details, see Fig.~6 in \citealt{wang+20}). 

Other studies using deep near-IR data combined with optical observations could help to disentangle the Milky Way foreground dwarfs from giants in NGC~5128, and thus to better quantify the fraction of the intermediate-age population. An example is shown in Appendix B for the major-axis field F5, where we find very few luminous AGB stars. 

When the NGC~5128 distance modulus of 27.91, the average reddening $E(B-V)=0.1$, and metallicity [Fe/H]$=-0.4$~dex are taken into account, the MS turnoff magnitude for a 2 Gyr population would have  I=19.26, J=29, H=28.9, and K=28.86 mag. A 4 Gyr population MS turnoff is fainter (I=30.76, J=30.45, H=30.2, and K=30.15). The bright MS turnoff for the intermediate-age population younger than $\sim 4$~Gyr will be within reach of the MICADO camera at the Extremely Large Telescope (ELT) throughout the halo of NGC~5128 beyond 1 $R_e$ \citep{schreiber+14, davies+21}.

\section{Stellar density profile and ellipticity of the halo}

In Paper I, we suggested based on a smaller dataset (only F1-F9, and with a less
thorough CMD cleaning procedure) that the outer halo of NGC 5128 might be more elongated than the inner halo, consistent with results from other large galaxies.
The present, more extended set of data can be used to carry out a more detailed
measurement of the structure of the diffuse halo. For this purpose, we used the star counts for the cleaned fields (Fig.~\ref{fig:cmdf1_9_clean} and \ref{fig:cmdW1_16_clean}).
RGB stars over all metallicities are included in this simple indicator, 
but no corrections are included for any intrinsic age differences of the stellar population as a function
of galactocentric distance.

\begin{figure}
\resizebox{\hsize}{!}{
\includegraphics{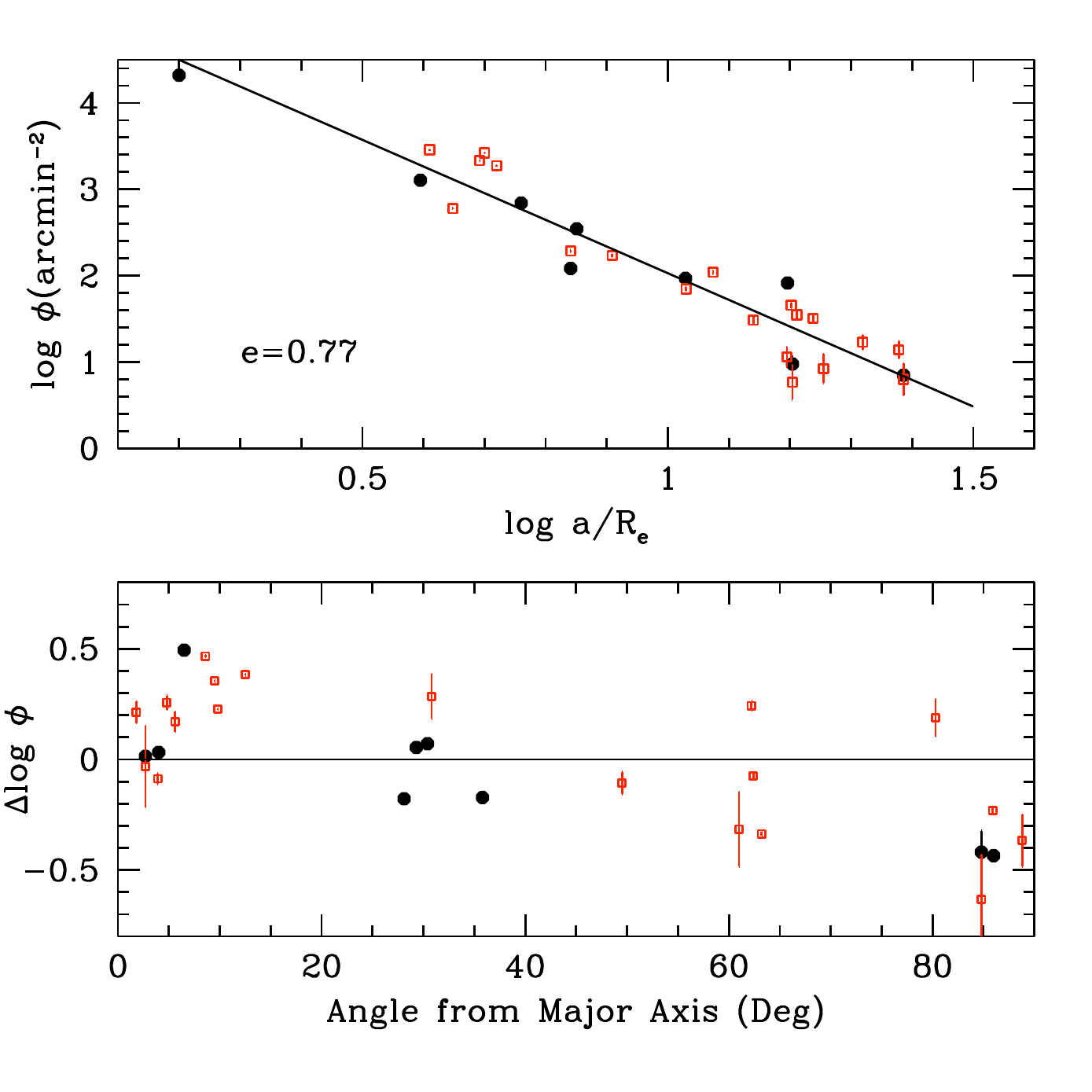}
}
\caption{\emph{} Number density of halo stars in NGC 5128 as a function of galactocentric
        distance $a$, normalized to the effective radius $R_e$ (upper panel).  This calculation assumes a halo elongation
        $e = 0.77$ from the inner-halo photometry of \citet{dufour+79}.  Black symbols are the ACS and WFPC2 fields from Table \ref{tab:HSTobslog},
     and open red symbols are the WFC3 fields. The equation for the power-law fit line is given in the text.  Error bars are calculated from Poisson statistics, including
     the raw number counts in each field, the number of subtracted stars from the decontamination procedure, and the completeness correction.\emph{}  Residuals from the fitted line in the upper panel, plotted vs. azimuthal angle 
from the isophotal major axis (lower panel). The residuals clearly show a systematic trend with position angle $\theta_a$,
indicating that an incorrect ellipticity $e$ has been assumed.}
\label{fig:density1}
\end{figure}

\begin{figure}
\resizebox{\hsize}{!}{
\includegraphics{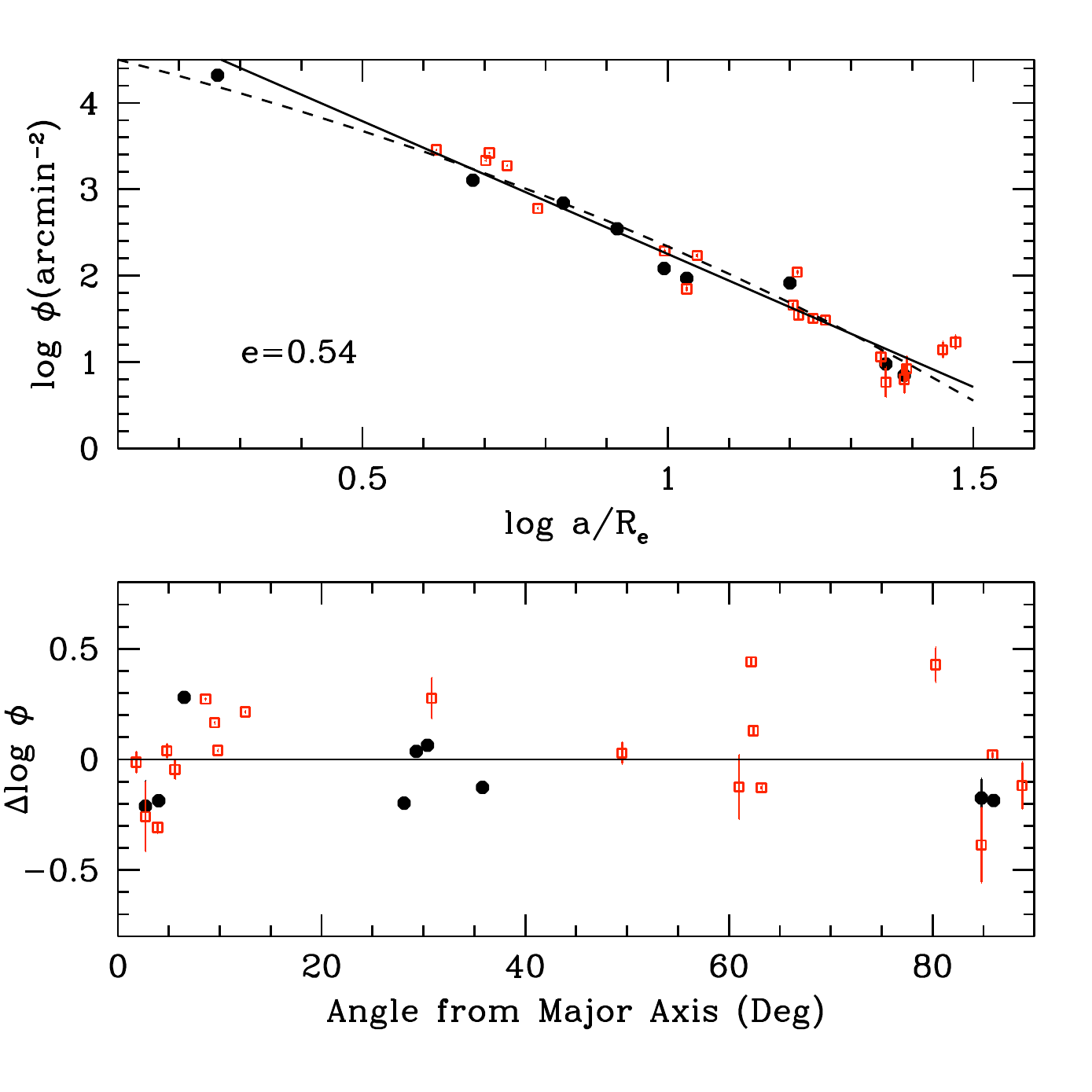}
}
\caption{Same data as in the previous figure, now recalculated with the
        assumption that the halo eccentricity is $e = 0.54$ (upper panel).
        The new fitted power-law line is given in the text.
        The \emph{dashed line} gives the best-fitting de Vaucouleurs $r^{1/4}$ 
        profile to the same data. Residuals from the power-law solution in the upper panel, plotted vs. azimuthal distance
from the isophotal major axis (lower panel).  Significant scatter remains, but there is no remaining trend with $\theta_a$.}
\label{fig:density2}
\end{figure}

To calculate the number density of RGB stars $\phi_{RGB}$ 
versus galactocentric distance, we used the
decontaminated (cleaned) CMDs and took all stars within the selection box $24 < I_0 < 26, 1.0 < (V-I)_0 < 1.4$.  The total numbers of stars within this region were explicitly corrected for
photometric incompleteness:  that is, each individual star was multiplied by $(1/f),$
where $f$ is the completeness factor at its given magnitude and color.
For the ACS and WFC3 fields F5-F9, the mean completeness correction
factor is just 1.1, while for the shorter-exposure fields W1-W16, the correction factor
averages about 1.34.  For F1-F3, which were taken with the less sensitive WFPC2 camera but
with relatively long exposures, the correction factor is about 1.06.  Finally, for the 
very deep F4, the correction is 1.00 over our magnitude range of interest.
Dividing by the area of the camera then gives $\phi_{RGB}$, now fully
corrected for contamination and incompleteness.

The location $(\alpha, \delta)$ of each field on the sky 
was converted into coordinates $(x, y)$ projected onto the major and minor axes of
the galaxy, assuming that the major axis is oriented 55\degree\ N of E (clockwise; see Fig.~1).  
From the basic equation of the ellipse and its eccentricity,
\begin{equation}
\frac{x^2}{a^2} + \frac{y^2}{b^2} = 1 \,\, {\rm and} \, \, b = ea
,\end{equation}
we obtain the necessary relation to convert any location $(x, y)$ into its equivalent
semimajor axis $a$ value,
\begin{equation}
a^2 = x^2 + \frac{y^2}{e^2}.
\end{equation}
For a first iteration, we assumed\emph{} that (a) the eccentricity $e$ is the same as the
value $e = 0.77$ for the inner halo isophotes (as used in Paper I); and (b) the orientation
angle of the major axis (55\degree N of E, i.e., clockwise) is a constant and valid at all radii.
However, because we have target fields over a wide area that reaches far beyond the inner halo
where $e$ was determined, we can also investigate if $e$ might change at large radii.

After calculating $a$(equiv) and $\phi_{RGB}$ for every target field, we plotted a first
estimate of the density profile.
The result is shown in Fig.~\ref{fig:density1}.  Here $a$ is normalized to the isophotal
effective radius adopted as $R_e = 305''$ (Paper I).

A least-squares fit with equal weights to all points in Fig.~\ref{fig:density1} gives
\begin{equation}
        {\rm log} ~ \phi_{RGB} \, = \, (5.117 \pm 0.201) - (3.088 \pm 0.193)~{\rm log}~(a/R_e)
,\end{equation}
where the units of $\phi$ are number per arcmin$^2$.  The residuals around this fit
have an rms scatter $\sigma({\rm log} \phi) = 0.290$ dex.  However, the residuals (shown in the lower panel, plotted versus azimuthal angle $\theta_a$) clearly
depend on position angle. 
That is, the fields closer to the \emph{\textup{minor}} axis ($\theta_a \rightarrow 90\degree$) 
have lower densities than expected under our assumptions.  An obvious interpretation is
that the equivalent $a-$values for these same fields should actually be higher than
we first assumed:  $e$ is even lower than we thought, and the halo is in fact more elongated at larger radii than it is in the inner halo.

We can therefore iterate the calculations of $a$ assuming different values of $e$, until a
result for the radial profile is achieved that leaves no systematic residuals in $\phi$ versus position angle $\theta_a$.  This same solution will also minimize the
field-to-field scatter around the best-fit curve.
We find that the solution converges at $e = 0.54$ with an estimated uncertainty $\pm 0.02$.
This solution is shown in Fig.~\ref{fig:density2}.
The new equation for the density profile in the same power-law form is
 \begin{equation}
{\rm log} ~ \phi_{RGB} \, = \, (5.328 \pm 0.153) - (3.077 \pm 0.138)~{\rm log}~(a/R_e) \, .
\end{equation}
The scatter around the best-fit curve is now minimized, at $\sigma({\rm log} \phi) = 0.212$ dex.

An alternative well-known form for halo light is the de Vaucouleurs-Sersic  profile
log $\phi = A + B (a/R_e)^{1/n}$ with $n=4$.  For the same assumption that $e = 0.54$ 
for the outer halo, the best-fit $r^{1/4}$ curve is also shown in Fig.~\ref{fig:density2}.
The result is
\begin{equation}
{\rm log} ~ \phi_{RGB} \, = \, (7.685 \pm 0.271) - (3.008 \pm 0.142) (a/R_e)^{1/4} \, .
\end{equation}
It is nearly as accurate as the simpler power-law form (the residual rms scatter is $\pm 0.224$, insignificantly different from the power law).

The residual scatter of the data points in Fig.~\ref{fig:density2} is clearly 
larger than expected from Poisson count statistics alone.  The scatter is
also noticeably the largest for the outermost fields, where the numbers of
counted RGB stars are particularly low.  This may be evidence that the halo becomes more patchy due to substructure at several $R_e$ and beyond, but
deeper photometry and larger area coverage are needed to pursue this interpretation further.

\begin{figure}
\resizebox{\hsize}{!}{
\includegraphics{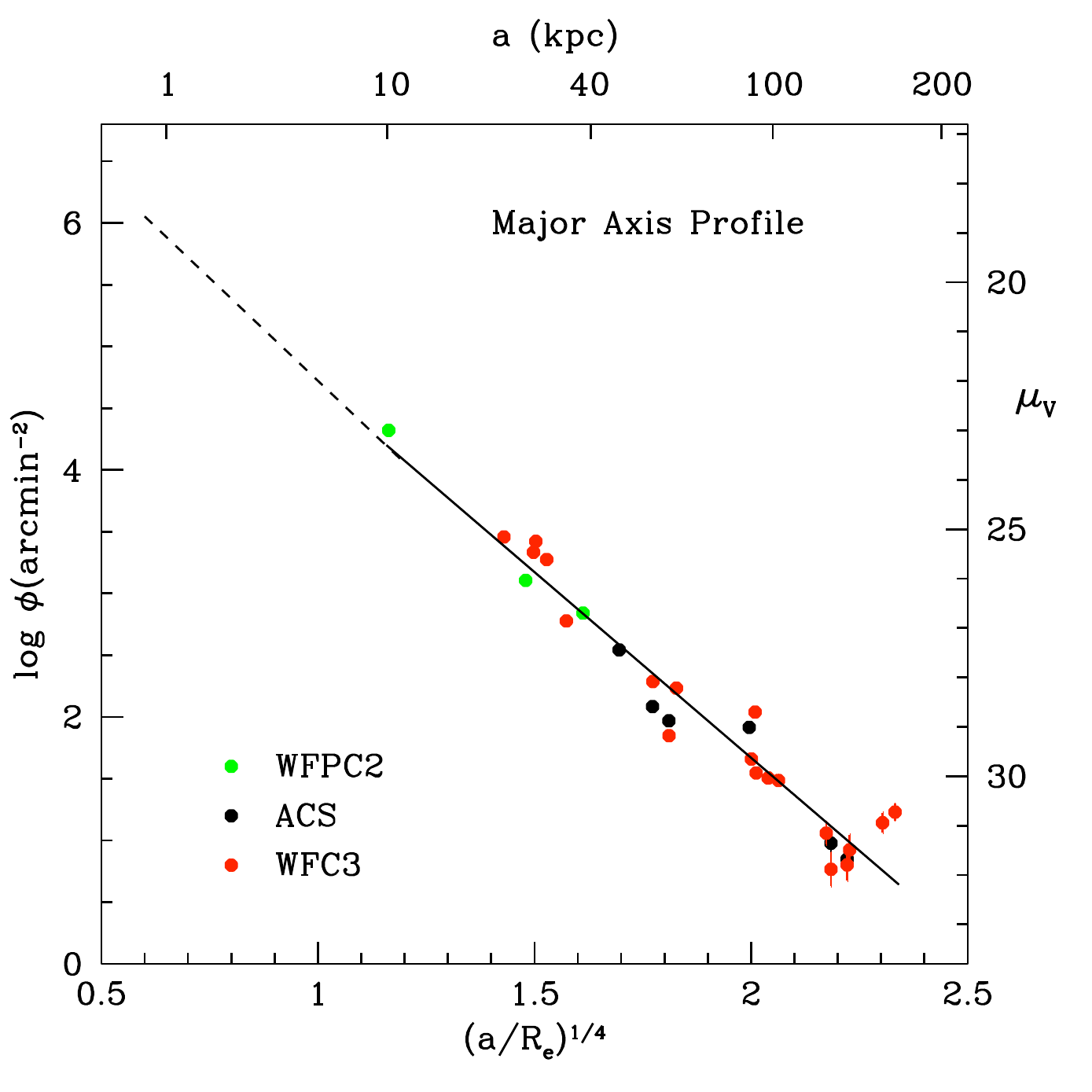}
}
\caption{Surface brightness profile along the isophotal major axis of NGC 5128 in the
form of the $a^{1/4}$ law, as described in the text.  Our star count data are shown as the
green, black, and red points from the three different HST cameras, along with the best-fitting
$a^{1/4}$ profile given in Fig.~\ref{fig:density2}, and they assume $e=0.54$ for the outer halo, 
as described above.  The \emph{dashed line} shows the inner-halo surface brightness profile
from \cite{dufour+79}, where $\mu_V$ is in mag arsec$^{-2}$.  The magnitude scale at the right
has been shifted vertically to match the inner surface brightness profile with the outer star counts at
$(a/R_e) = 1.158$.}
\label{fig:profile}
\end{figure}

Last, for this section, we show in Fig.~\ref{fig:profile} the starcount data obtained here
combined with the surface brightness profile for the inner part of the galaxy as obtained
by \cite{dufour+79}, which is
$\mu_V = 8.32 [(a/R_e)^{1/4} - 1] + 22.00$ mag arcsec$^{-2}$.  The purpose of this plot is simply to give a comprehensive radial profile
of the galaxy along its major axis, extending from $a \simeq 0.23 R_e$ out to almost $30 R_e$.
Our very innermost HST field F1 just barely overlaps the \emph{\textup{outermost}} limit of the direct
surface brightness photometry \citep{dufour+79}, so splicing the two sets of data together 
is intrinsically uncertain.  Nevertheless, they can be successfully matched for a conversion offset where log $\phi = 0.0$ (1 RGB star per arcmin$^{-2}$ between $I = 24 - 26$) is equivalent to a surface brightness $\mu_V \simeq 33.82$ mag arcsec$^{-2}$ \footnote{For a stellar population model based estimate of stellar density within 2 mag of the TRGB vs. surface brightness see {\tt https://eso.org/sci/meetings/2015/StellarHalos2015/ \\ talks\_presentation/greggio\_stellarhalo.pdf}}.

As Fig.~\ref{fig:profile} shows, our RGB star-count data in the most remote
parts of the halo reach $\mu_V \sim 32$ mag arcsec$^{-2}$ (roughly
equivalent to $\mu_B \simeq 33$; see the figure in Appendix C).  These results exemplify the ability of
resolved-star photometry to penetrate to surface brightness levels that can scarcely
be obtained by other means.

Early evidence that the NGC 5128 halo has higher 
ellipticity  at larger galactocentric distance 
was provided by \citet{hesser+84} on the
basis of the spatial distribution of the halo 
GCs
and on the isophotal contours
of the outer halo from
deep Schmidt-telescope images that were available then \citep{cannon1981}.
More recently, \citet{crnojevic+13} used VLT/VIMOS ground-based imaging of selected fields along the
major and minor axes.  Compared with our HST fields, their data have the advantage of covering a larger total area, 
but the disadvantage of much higher field contamination.  
Their study traces the NGC 5128 halo out to 85 kpc (slightly more than half as
far as our current study) and finds that the data match an outward extension of the
de Vaucouleurs profile established in the
inner regions \citep{dufour+79} moderately well.  Interestingly, they also suggest that
for $R_{gc} \gtrsim 55$ kpc, the halo shows evidence of becoming more elliptical, in the range 
$e \sim 0.5-0.6$.  Their estimate is very much in line with our finding of
$e = 0.54 \pm 0.02$ for $R_{gc} \gtrsim 30$ kpc.  

A systematic increase of stellar halo ellipticity
with radius is also a commonly observed feature of large ETGs \citep[e.g.,][]{tal+vandokkum11,huang+2018}. The \citet{cooper+13} simulations reproduced the observed stellar density profiles well. These simulations combined the semianalytic galaxy formation model of \citet{guo+11}, based on halo merger trees  from Millennium II N-body simulation \citep{boylan-kolchin+09}, with the particle-tagging technique of \citet{cooper+10} 
to resolve
low surface brightness outskirts of galaxies on scales from Milky Way-like galaxies to central galaxies of groups and poor clusters. The 2D images of simulated galaxies, scaled appropriately to display also the low surface brightness features, have quite elongated stellar halos that reach $\mu_V \ga 32$~mag/arcsec$^2$ at $150-200$~kpc projected galactocentric distance for galaxies in the mass range of NGC~5128 ($M_\star \sim 2 \times 10^{11}$~M$_\sun$; \citealt{fall+romanowski18}). The halo surface brightness and stellar mass density profiles of simulated galaxies are (by construction) best fit by the sum of two Sersic profiles. A single Sersic profile is only a good fit for the circularly averaged total stellar surface density for galaxy halos in which the accreted component dominates at all radii \citep{cooper+13}. The accretion-dominated galaxies have more extended profiles and a higher Sersic index than in situ dominated galaxies, but even for these galaxies, the in situ stars are an important component within the inner $\la 10$~kpc. In a recent study, \citet{pulsoni+21} examined the impact of galaxy merger history on the intrinsic shape profiles of early-type galaxies in the TNG100 Illustris simulation \citep{nelson+19}. They found that galaxies with higher stellar mass and accreted fractions are less flattened and more triaxial, and that mergers contribute to more spherical-triaxial stellar halo shapes.  

A ground-based imaging study with the Magellan Megacam
covers a still larger and contiguous area out to $R_{gc} \simeq 150$ kpc \citep[][part of the PISCeS survey]{crnojevic+16}.  Their study 
reported two low-surface-brightness streams in the outer halo,
as well as faint satellites.  The HST fields in our current study typically avoid
these satellites or tidal features and thus are closer to measuring the ``smooth halo''
of the galaxy (as far as such a smooth underlying distribution exists in these
outer regions). However, we note that several of our pointings that are close on the sky show larger scatter, and the most notable exception to the smooth-halo field is pointing F6A, which is traversed by the disrupting dw3 dwarf stream  \citep{crnojevic+19}.

\section{Summary and conclusions}

We have analyzed HST imaging of the halo of NGC 5128 to study the
radial density profile of its halo, covering the distance range $R_{gc} = 8$ to 140 kpc
(equivalent to 1.5 to 25 $R_e$).  Data from the WFPC2, ACS, and WFC3 cameras
for a total of 29 distinct pointings across the smooth halo were used,
and one of these pointings acted as a control field for background
subtraction. These
target fields span a diameter of 4\degree \ on the sky.
The projected number density of red giant halo stars in the magnitude range
$I_0 = 24 - 26$ over all metallicities was used as the halo tracer after suitable
subtraction of background field contamination and correction for photometric
incompleteness.

\begin{itemize}
    \item For magnitudes $I_0 < 24$, we find that the dominant source of field
    contamination is from Milky Way foreground stars that are brighter 
    than the NGC 5128 RGB population, as determined either from a control field
    or from TRILEGAL models.
    \item In the most remote fields along either the major or minor axes 
    of the halo, we were able to successfully trace the RGB population
    down to 
     an equivalent surface brightness $\mu_V \simeq 32$ mag arcsec$^{-2}$.  With wider area coverage and/or deeper data,
    still fainter equivalent surface brightness levels could be achieved. The Nancy Grace Roman Space Telescope will be an ideal instrument for mapping extended nearby galaxy halos \citep{williams+19}.
    \item Over the full radial range covered by our study, the profile is well
    matched by a classic $r^{1/4}$ curve or (within the scatter of the data)
    a simple power-law form $\phi \sim r^{-3.1}$.  Despite the fairly steep
    decline of the profile, we have not yet discovered any final
    cutoff or ``end'' of the halo of this giant galaxy, even with the best 
    available data. We detect some substructures shown by the scatter in stellar surface density around the best-fit profile, which agrees with the contiguous panoramic imaging by the PISCeS survey \citep{crnojevic+16}. However, most of our fields avoid obvious substructures and measure the ``smooth halo'', as much as such a smooth component exists in the outskirts of a large galaxy.
    \item Classic isophotal measurements showed that NGC 5128 has an inner-halo 
    ellipticity $e= (b/a) = 0.77$.  However, our data show that for $R_{gc} \gtrsim 30$ kpc, the halo flattens to $e = 0.54 \pm 0.02$.
    \item The analysis of the luminosity function shows that
    over $22 < I_0 < 23.86$, that is, above the TRGB, a measurable excess of stars above the background level predicted by the Milky Way population models (both TRILEGAL and Besan\c{c}on) is 
    present. Beyond 30 kpc, this excess is constant and consistent with an additional Milky Way dwarf-star population that is missing from the models, as is evident also from the combination of optical and near-IR photometry (see Appendix B). At galactocentric distances smaller than 30 kpc, a 2-3 Gyr old AGB component contributes between 10-40\% of the stellar mass. This confirms earlier results from LPVs, GCs, and red clump stars \citep{rejkuba+03_LPVnir, woodley+10a, rejkuba+11}. Further studies combining optical and near-IR data can provide powerful diagnostics distinguishing the Milky Way dwarfs from AGB giants in NGC 5128, and the ELT will be able to resolve the turnoff for this intermediate-age population.
\end{itemize}

The observed stellar density profile, shape, and substructures in the stellar halo of NGC~5128 testify to its growth through hierarchical mergers and show the dominant contribution from accreted stars in its outskirts. It does not  answer the question whether the galaxy formed through a relatively recent ($<5-6$~Gyr) major merger of two spiral galaxies \citep{bekki+peng06, wang+20} or had a two-phase formation \citep{oser+10} with an early assembly stage that formed the elliptical, which subsequently grew through minor-merger episodes and accretion of smaller satellites \citep{naab+09}. In its inner parts, NGC~5128 shows clear evidence of a past merger history, and our study uncovered a quite extended distribution of $2-3$~Gyr old intermediate-age stellar population. The accretion of a small gas-rich spiral may have provided the fuel for the recent star formation in the central parts as suggested by e.g., \citet{quillen+93}. However, the short time since the merger in this model ($< 2 \times 10^8$ yr) would imply that the bright intermediate-age component we observe within the inner 30~kpc was deposited directly from the accreted galaxy or originated from a different merger event. In a following paper, we will use this dataset to derive
the metallicity distribution function of the RGB stars along with their
progressive change with radius.

\begin{acknowledgements}
We thank the referee for the thorough review of the paper.
This research has made use of the NASA/IPAC Extragalactic Database (NED) which is operated by the Jet Propulsion Laboratory, California Institute of Technology, under contract with the National Aeronautics and Space Administration. This research is based on observations made with the NASA/ESA Hubble Space Telescope obtained from the Space Telescope Science Institute, which is operated by the Association of Universities for Research in Astronomy, Inc., under NASA contract NAS 5–26555. 
The HST observations are associated with programs 5905, 8195, 9373, 12964, 13856, and 15426.
This paper also used data collected at the European Southern Observatory under ESO programme 290.B-5040(A) and 290.B-5040(B).
Research by DC is supported by NSF grant AST-1814208, and by NASA through grants number HST-GO-15426.007-A and HST-GO-15332.004-A from the Space Telescope Science Institute, which is operated by AURA, Inc., under NASA contract NAS 5-26555.
\end{acknowledgements}


\bibliographystyle{aa}
\bibliography{N5128_HST}


\begin{appendix}

\section{Luminosity and color functions of Milky Way models: Comparison to data}

\begin{figure}
\centering
\resizebox{\hsize}{!}{
\includegraphics[angle=0,clip]{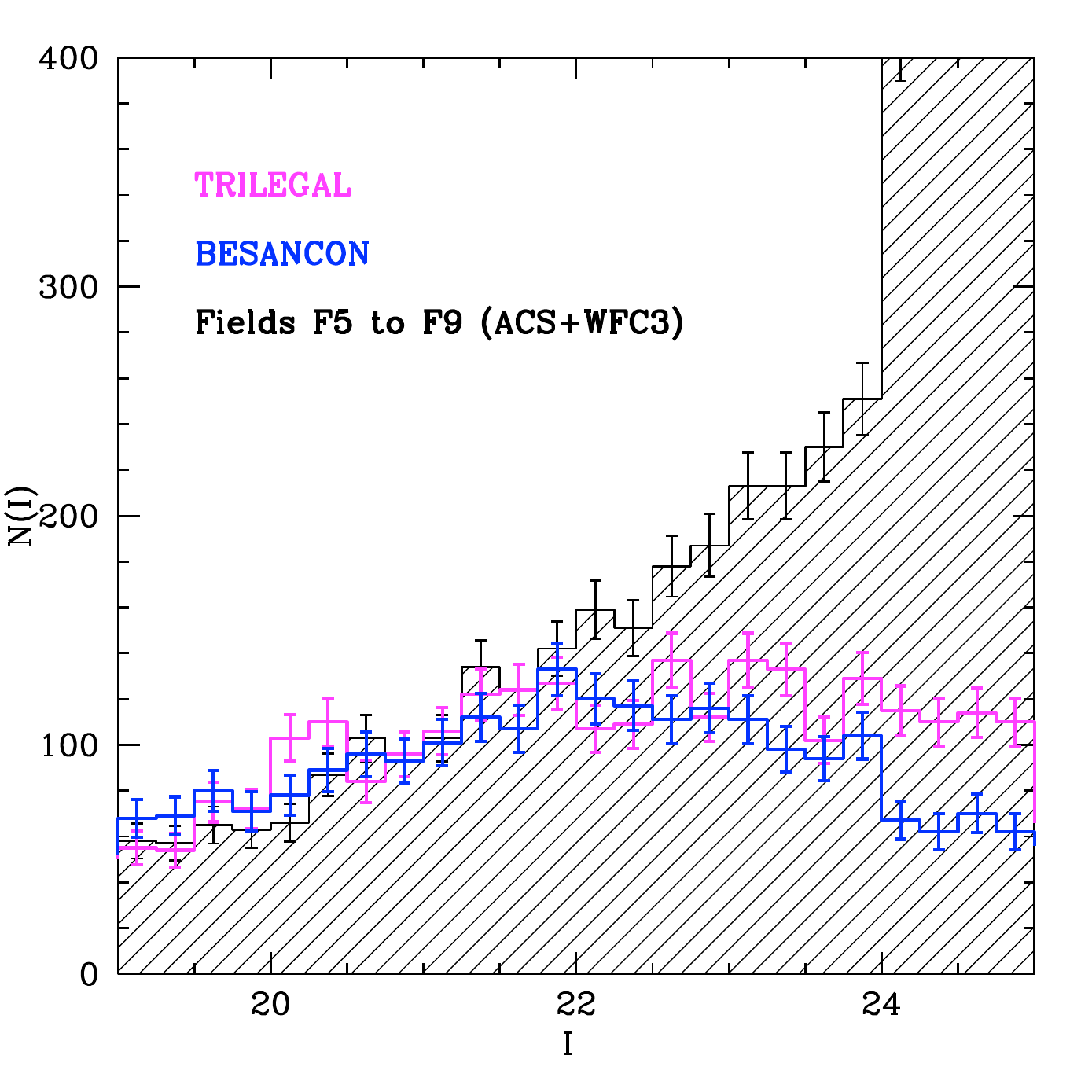}
}
\caption{Comparison of the global luminosity function of Cycle 20 fields (shaded area) with the Milky Way models from the TRILEGAL (magenta) and Besan\c{c}on (blue) simulators.} 
\label{fig:cena_lfs}
\end{figure}

\begin{figure}
\centering
\resizebox{\hsize}{!}{
\includegraphics[angle=0,clip]{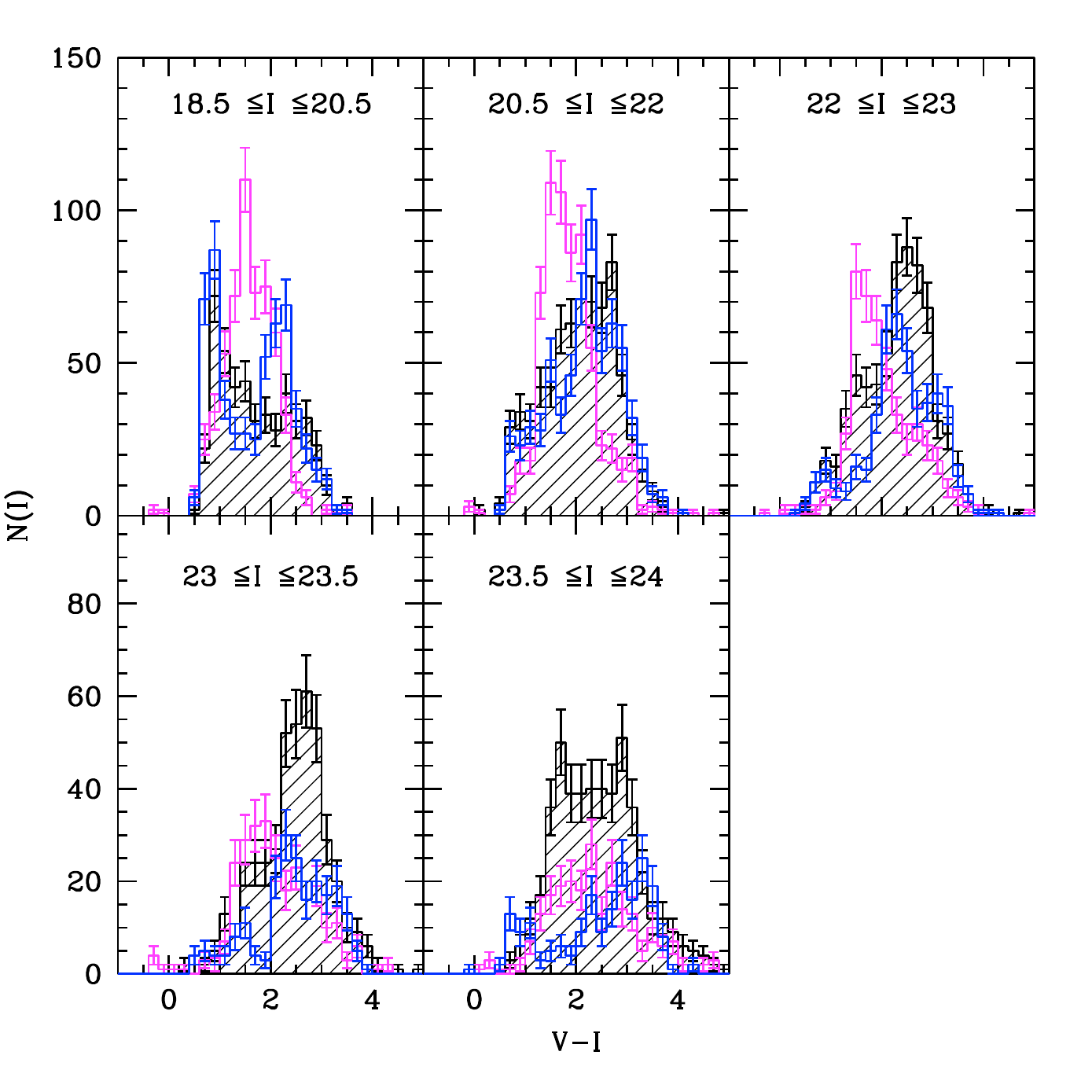}
}
\caption{Comparison of the observed and model color functions in different magnitude bins. The coding is the same as in Fig. \ref{fig:cena_lfs}}
\label{fig:cena_cfs}
\end{figure}

The comparison of the foreground simulations to the luminosity and color function
of the bright stars in our combined five-field CMD shows similar features for TRILEGAL and Besan\c{c}on simulations (see Fig. \ref{fig:cena_lfs} and \ref{fig:cena_cfs}):
\begin{enumerate}
\item At $I \leq 22,$ the LF is very well reproduced with both simulators. The color functions are better reproduced with the Besan\c{c}on model because the TRILEGAL model shows a too narrow plume of stars with too blue median color. 
\item In the $22<I<23$ range, the data show an excess of stars with respect to both simulations (see Fig.~\ref{fig:cena_lfs}), which could either be foreground stars or members of NGC 5128. In this bin, the color of the TRILEGAL simulated stars peaks to the blue of the observed distribution, while the Besan\c{c}on model agrees better with the data, even though it underrepresents the data at $V-I \sim 2.75$.
\item In the $23<I<24$ range the width of the color distribution is quite well reproduced by both models, but not the counts. Interestingly, in $23.5<I<24,$ the TRILEGAL model shows two peaks located at the same colors as in the data.
\end{enumerate}

The nature of these excess stars can be investigated by adding near-IR information to the optical CMD.

\section{Near-IR photometry: Foreground contamination, or AGB contribution?}

The combination of the near-IR and optical photometry is a useful tool for assessing the level of foreground contamination as well as the potential presence of AGB stars that are particularly bright in near-IR bands and are located above the tip of the RGB. We took advantage of the last "delta" call for observations with ISAAC before the instrument was decommissioned to secure near-IR $JK_s$ imaging for fields scheduled on the HST WFC3 camera during Cycle 20. ISAAC was the near-IR imager and spectrograph on the ESO VLT, at that time, mounted on the UT3 Melipal. We used the so-called short-wavelength mode of ISAAC that offered the $1k \times 1k$ Hawaii detector with 0.148 arcsec/pix resolution. Observations were carried out with $K_s$ and $J_s$ filters, which were recommended for accurate stellar photometry. From here on, we drop the subscript $s$ and shorten the filter names to $J$ and $K$.
Detector integration time (DIT) was set to 12 sec and the number of DITs (NDITs) to 8 for K band.  For J band, DIT was 34 sec and NDIT 5. The ISAAC detector averages NDIT observations each with DIT s exposure, and therefore each individual K-band image had 12 s average exposures and J-band images had 34 s exposure averages.  All observations were carried out with the so-called auto-jitter template, that is, offsetting within the same field after the previous (DIT$\times$NDIT) image was recorded.

The field of view of ISAAC matched the size of the WFC3 field of view very well, but at a much lower resolution (factor 3.7). Unfortunately, when the ISAAC observations were scheduled, the HST program had been only partially executed, and given the free position angle for the planned HST observations, which was selected to maximize the scheduling of the observations, this meant that we could not fully align the near-IR ISAAC and optical WFC3 images.  Some of the fields therefore overlapped only partially, and because the instrument was decommissioned before all observations could be completed, we have good-quality data for both filters with full WFC3 field overlap for only F5W and F9W. Given the very low stellar density in F9W, we compare the near-IR and optical data for F5W in detail below.

The ISAAC data reduction followed the steps described in \citet{rejkuba+03_LPVcat}: dark subtraction, flat-fielding, and two-step combination with a rejection of bright pixels associated with stars or other detected objects. The last step was carried out with the IRAF task \emph{xdimsum} and a series of custom-made scripts. The photometry was carried out with \emph{daophot,} and the photometric calibration is based on the average zeropoint offset per filter computed from 8 to 16 high-quality 2MASS stars (all brighter than $K_\mathrm{2MASS}<15.7$) matched with well-measured stars in ISAAC images in each field. 

\begin{figure}
\centering
\resizebox{\hsize}{!}{
\includegraphics[angle=0,clip]{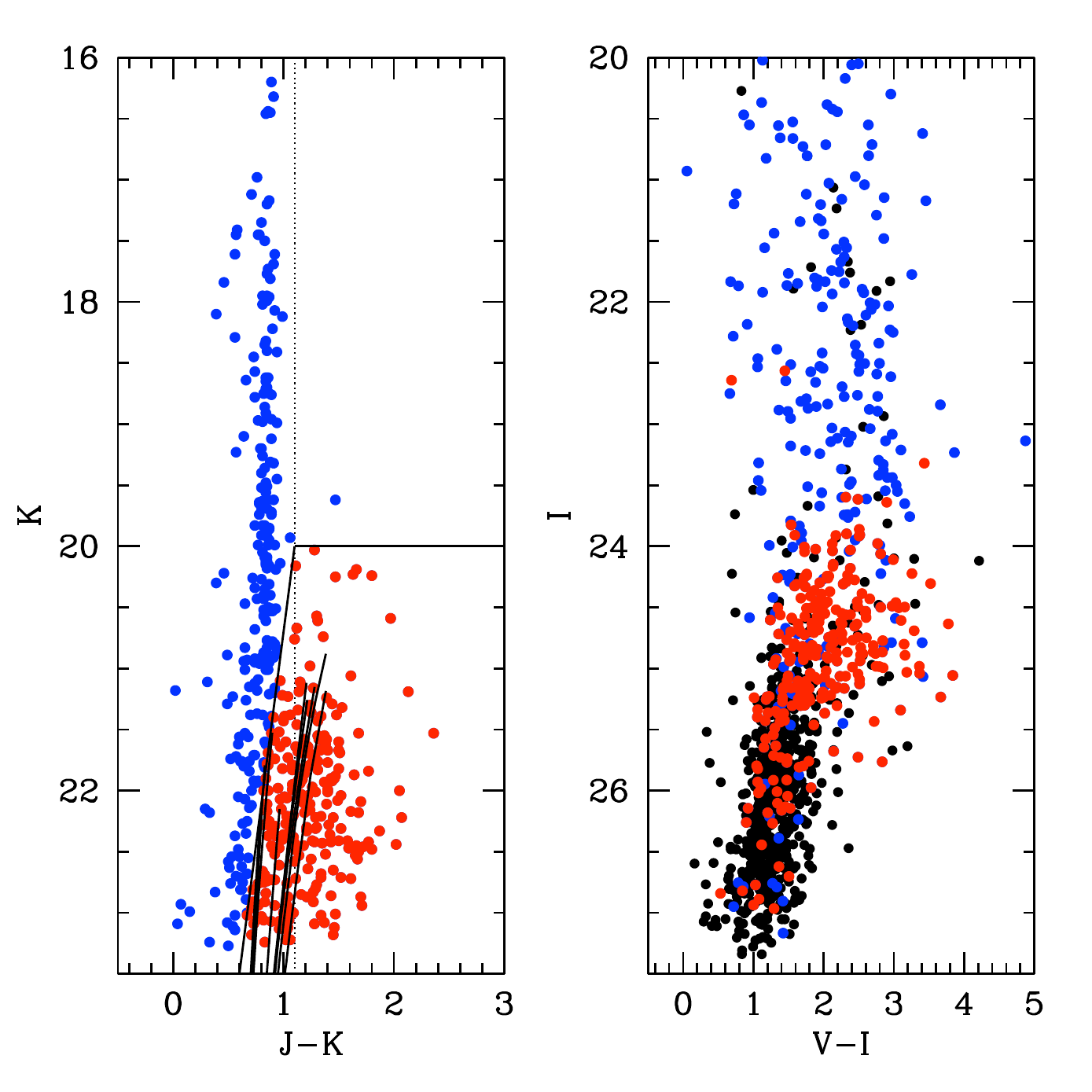}
}
\caption{\emph{Left:} Field F5 J-K vs K CMD based on ISAAC aperture photometry with sources selected from the HST image. Overplotted are RGB fiducials for GCs
from \citet{ferraro+00}. \emph{Right:} HST CMD for F5W on which stars detected also in ISAAC images are shown with blue (foreground) and red (likely RGB members) symbols according to selection shown in the left pannel.
}
\label{fig:ISAAC_HSTphotometrycomp}
\end{figure}

Point spread function fitting photometry with \emph{allstar} based on the J- and K-band images alone reaches the TRGB, but given the deeper HST photometry, we were also able to detect stars that are within the background noise on individual ISAAC images. To do so, we registered the HST catalogs to the ISAAC reference frame and used the position of stars that are well measured in HST images as input for the aperture photometry from ISAAC. This results in gaining about one magnitude in photometric depth for the faintest measured stars in near-IR images (Fig.~\ref{fig:ISAAC_HSTphotometrycomp}).

\begin{figure}
\centering
\resizebox{\hsize}{!}{
\includegraphics[angle=0,clip]{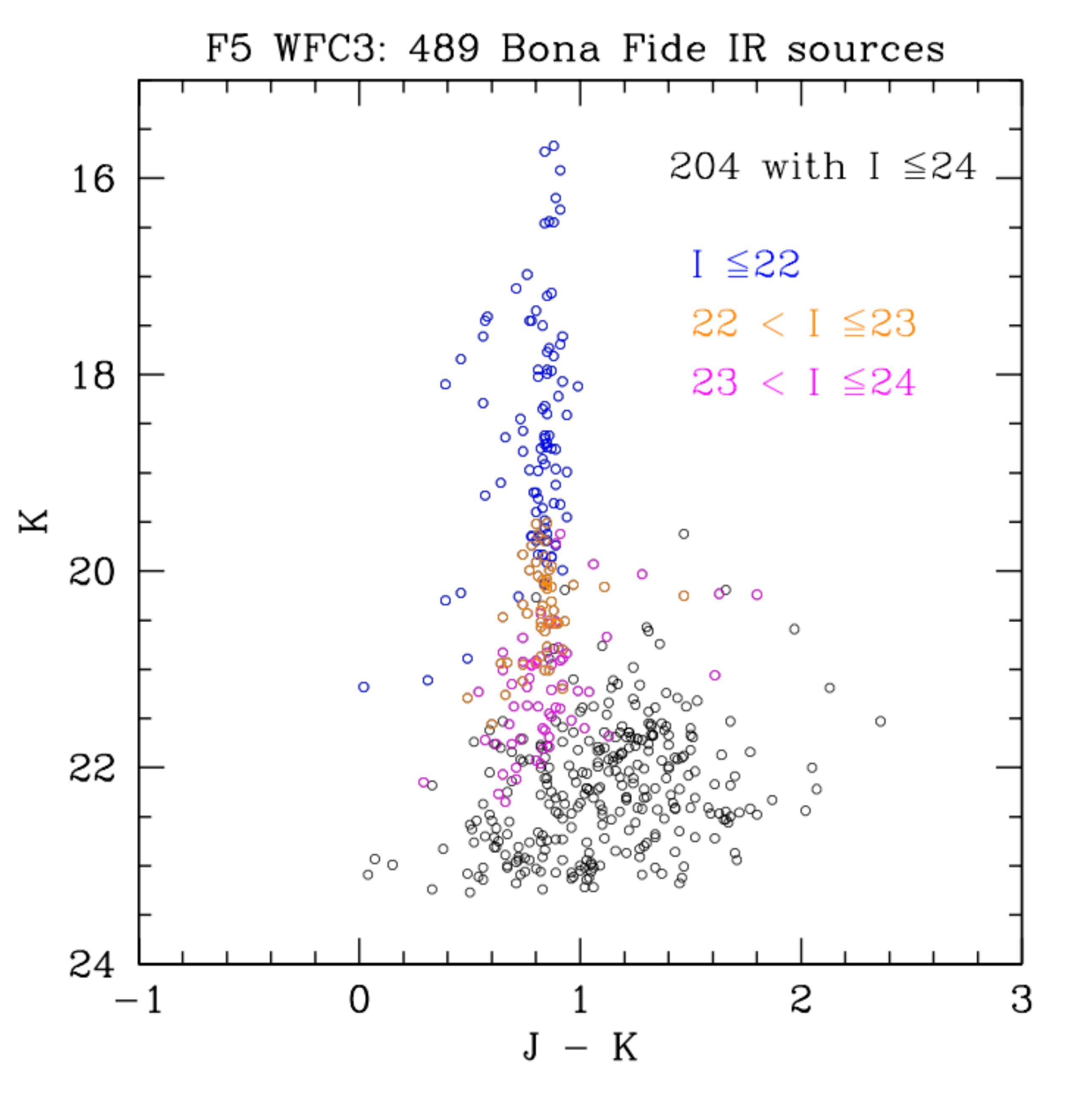}
}
\caption{Infrared CMD of stars detected in the ISAAC and WFC3 
pointings in Field 5. There are 489 sources with photometric error smaller than 0.5 mag in all V, I, J, and K bands. The color encodes the stellar magnitude in the I 
band. Colored points fall above the TRGB. 
}
\label{fig:ir_cmd}
\end{figure}

\begin{figure}
\centering
\resizebox{\hsize}{!}{
\includegraphics[angle=0,clip]{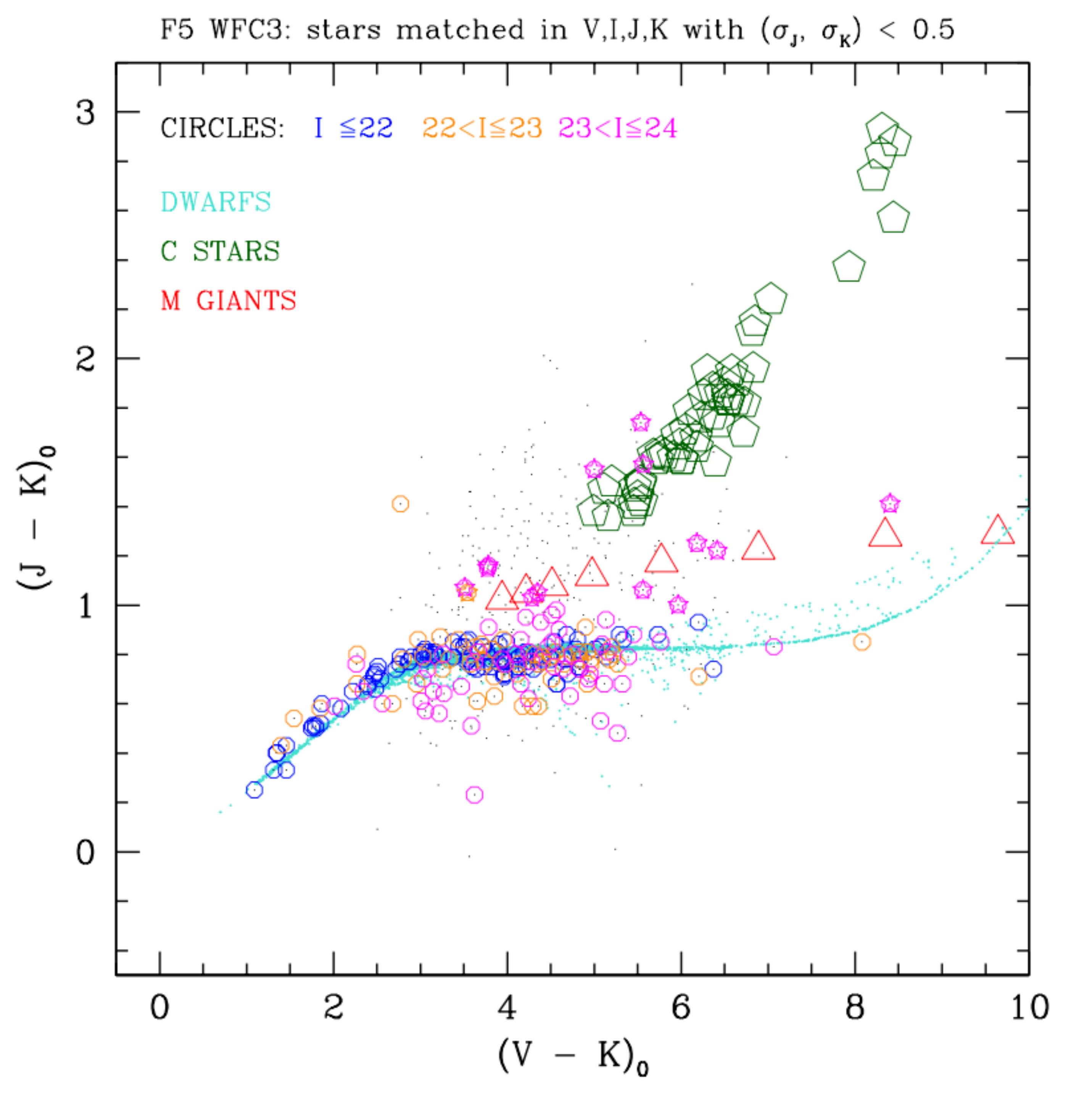}
}
\caption{Two-color diagram of the stars observed in F5W (gray dots) compared to the expected colors for stars of different type: dwarfs from the TRILEGAL simulations (cyan dots), M giants from \citet{fluks+94} empirical data (large red triangles), and
carbon stars (green pentagons) from \citet{bergeat+01}. The circled dots are stars with different I magnitude, as labeled and also indicated in Fig.~\ref{fig:ir_cmd}. Sources that are further highlighted with magenta stars are observed NGC~5128 AGB giants with magnitudes brighter than TRGB. 
}
\label{fig:two_cols}
\end{figure}

There are 1819 sources detected in the optical F5W HST data, 600 of which had near-IR magnitudes measured on ISAAC data. Selecting IR sources with  $\sigma \leq 0.5$ 
in J and in K leaves 489 stars with $VIJK$ photometry in this field.
Figure~\ref{fig:ir_cmd} displays the IR CMD. Bright stars, those located above the RGB tip, are color-coded according to their $I$-band magnitude\footnote{This field has an average reddening (see Tab.~\ref{tab:HSTobslog}) resulting in $A_I=0.168$ and observed TRGB at $I=24.03$.}. 
The great majority of $I \leq 24$ stars belong to the vertical feature at $J-K \sim 0.9$, which is due to foreground stars. This is especially true for stars brighter than $23$, but also for many
of the stars in the $I=23-24$ bin. Stars with $22<I<24$ and colors redder than $J-K>0.9$ could be members of NGC~5128. 

The diagnostic plot in Figure~\ref{fig:two_cols} is used primarily to distinguish between bright (AGB) giants and foreground dwarf stars: all matched stars (gray dots) are plotted in the two-color diagram.  Orange circles highlight stars brighter than $I=23$,
and purple circles highlight stars in the range $24 < I < 23$, that is, within one magnitude brighter than the TRGB. 
Expected colors from models and from empirical scales are overplotted in cyan (dwarfs), orange (giants), and green (carbon stars). Clearly, most of the matched stars brighter than the TRGB are dwarfs, clustering around the cyan dots, but a few are found in the giant region and some even on the carbon star sequence. 
The gray dots without circles have $I\geq 24$, consistent with being giants, members of NGC~5128. Their distribution is not particularly clustered around the giant loci due to the very large photometric errors of the near-IR measurements. We verified that stars with near-IR measurement errors $<0.3$ mag show  consistency with giant star colors. 

\begin{figure}
\centering
\resizebox{\hsize}{!}{
\includegraphics[angle=0,clip]{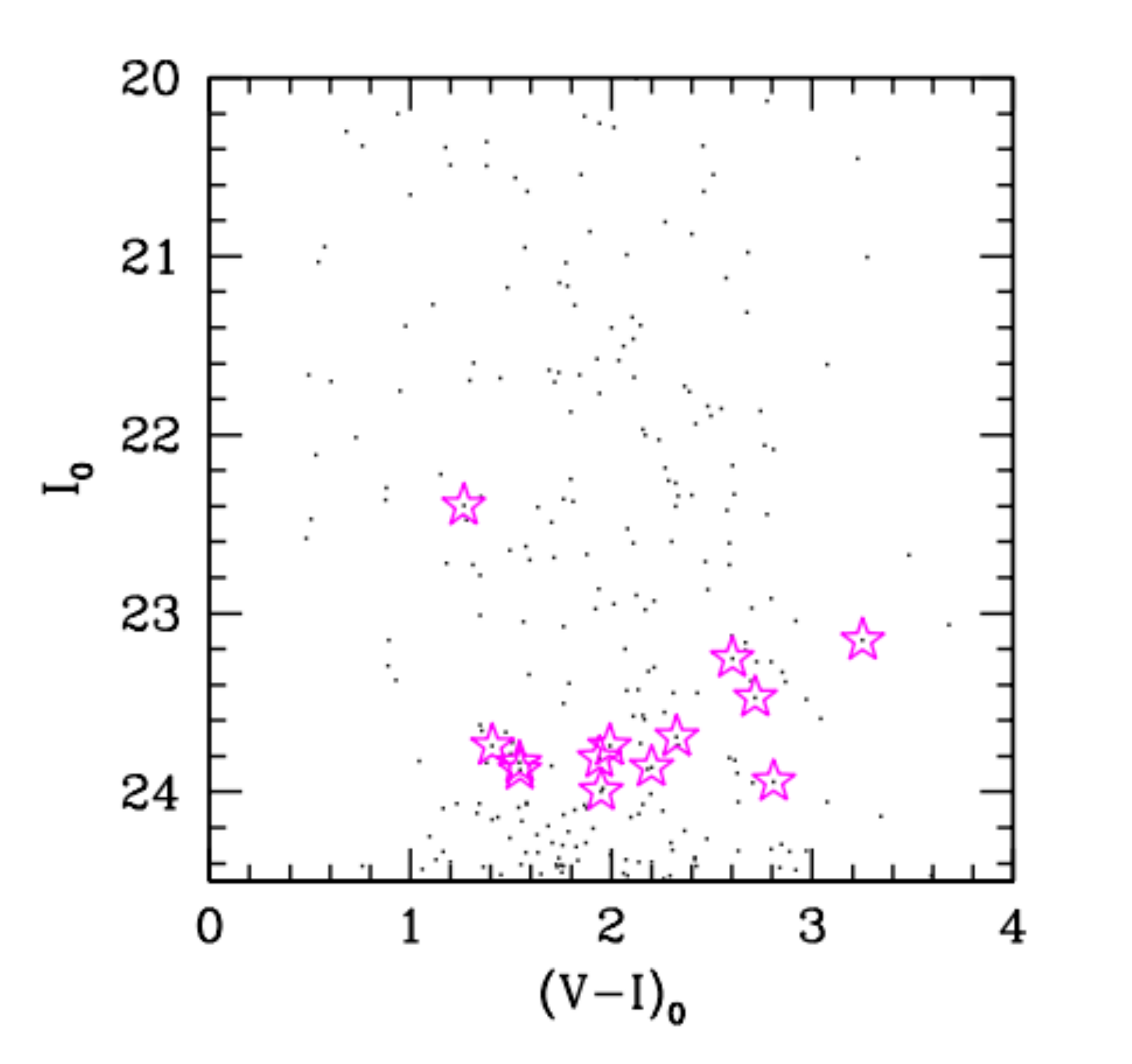}
\includegraphics[angle=0,clip]{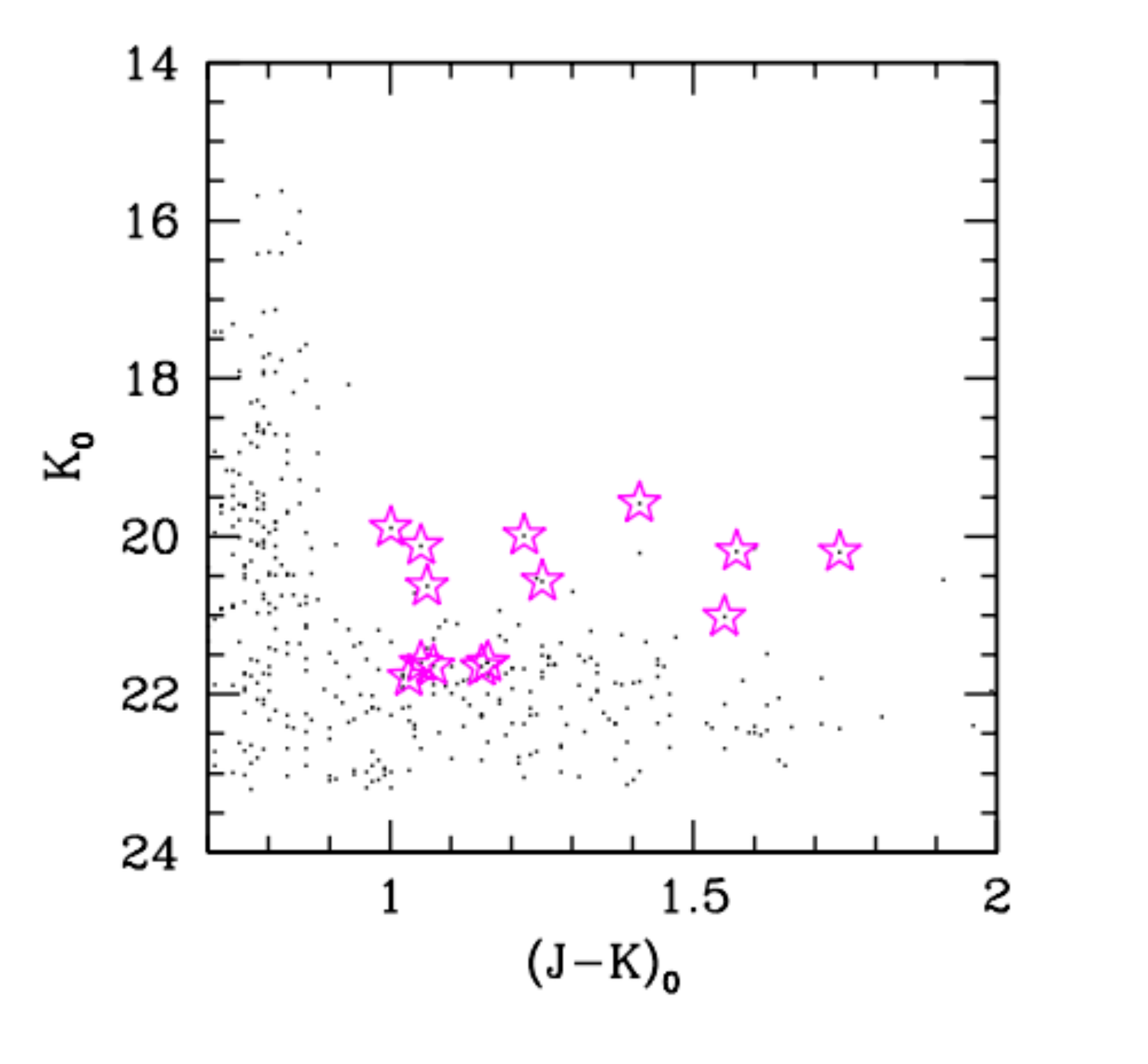}
}
\caption{Candidate members of NGC 5128 with magnitudes brighter than the TRGB are highlighted in the optical (left) and IR (right) CMD.  
}
\label{fig:wru}
\end{figure}

Fig. \ref{fig:two_cols} shows 13 stars with $J-K>1$ and brighter than $I=24$. These are highlighted in Fig.~\ref{fig:two_cols} as magenta stars and are plotted with the same symbol in Fig. \ref{fig:wru}, where we show their position in the optical and infrared CMDs. All but one of these stars are red in $V-K$ and in $V-I$ as well, and 11 of 
them lie just above the TRGB in the optical CMD. We conclude that these are bona fide AGB star members of NGC 5128 in field F5W, which is located at $\sim$ 10 R$_e$. This conclusion is supported by the position of these stars in the IR CMD. In addition, the three reddest stars in $J-K$ appear to be very good carbon-star candidates.

In summary, the great majority of the stars in our CMDs at magnitudes brighter than $I=24$ are  foreground dwarfs. The TRILEGAL and the Besan\c{c}on Milky Way models account very well for the foreground population at magnitudes brighter than $I=22$, and both slightly underestimate this component in the range $22 \leq I \leq 24$. In this range, we detect a few AGB members of NGC~5128, but most of the stars are likely Milky Way foreground dwarfs. 

\section{Surface brightness profile from star counts and integrated-light photometry}

\begin{figure}
\resizebox{\hsize}{!}{
\includegraphics{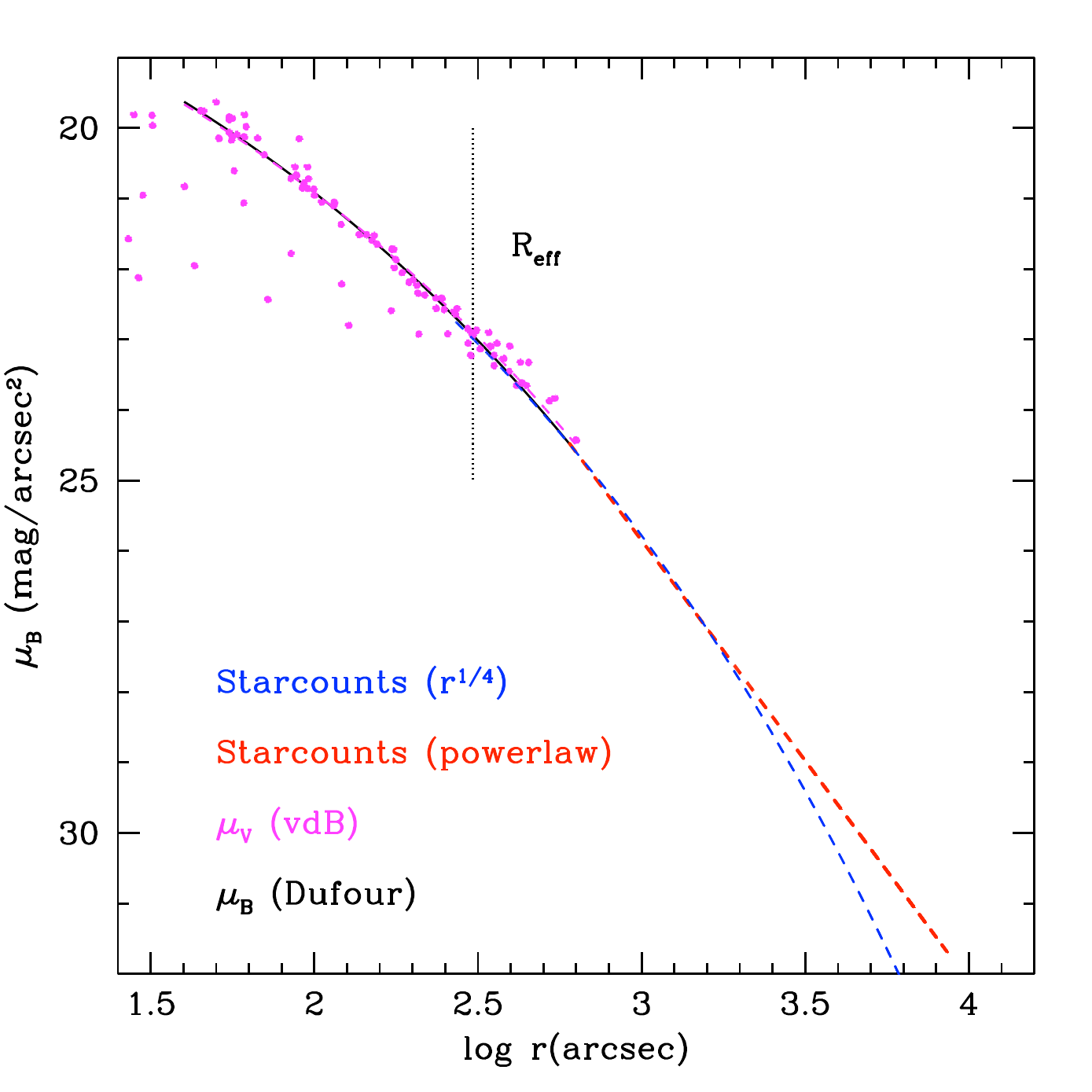}
}
\caption{NGC 5128 surface brightness profiles from \citet[][magenta line and dots, converted into $\mu_B$]{vandenbergh76} 
and \citet[][solid black line]{dufour+79} extended outward using the fits to the 
star counts from Paper I.  The red line is from the power-law fit
to the star counts, while the blue line is from a de Vaucouleurs
$r^{1/4}$ law fit to the starcounts.
} 
\label{fig:sb_profile}
\end{figure}

The surface brightness profile for the NGC 5128 halo as derived in Paper I is shown in Fig.~\ref{fig:sb_profile}.  
The inner parts are determined from the photometric integrated-light measurements by 
\citet[][magenta dots]{vandenbergh76}, with profile fits from 
\citet[][magenta line]{vandenbergh76} and \citet[][solid black line]{dufour+79}
derived from data that extend slightly farther than 1 $R_e = 305''$.  This integrated light photometry is spliced to a surface brightness profile based on  
star counts for halo RGB stars in the range $24 < I_0 < 26$ that extend the photometry outward (see the equations from Fig. 5 of Paper I). 
The minimum lower curve in blue is the $r^{1/4}$ law solution to our star counts,\begin{equation}
\textrm{log}~\phi = 7.7 - 2.1~r^{1/4}
,\end{equation} 
while the maximum upper curve in red is the power-law solution to the same star counts,
\begin{equation}
\textrm{log}~\phi = 6.6 - 2.6~\textrm{log}~r
,\end{equation}
where $\phi$ is the number per arcmin$^2$ and $r$ is the projected
galactocentric distance in kiloparsec.

\end{appendix}

\end{document}